\documentclass[aps,prl,reprint]{revtex4-2}
\usepackage{blindtext}
\usepackage{graphicx}
\usepackage{subcaption}
\usepackage{xcolor}
\usepackage[fleqn]{amsmath}
\usepackage[font=small,skip=3pt]{caption}
\usepackage{makecell}
\usepackage{hyperref} 
\usepackage{cleveref}
\usepackage{url}
\begin{document}
\title{Turbulent Transport-Limited Pedestals in Tokamaks}
\author{J. F. Parisi$^{1}$}
\author{D. R. Hatch$^{2,3}$}
\author{P. Y. Li$^{2}$}
\author{J. W. Berkery$^{1}$}
\author{A. O. Nelson$^{4}$}
\author{S. M. Kaye$^{1}$}
\author{K. Imada$^{5}$}
\author{M. Lampert$^{1}$}
\email{jparisi@pppl.gov}
\affiliation{$^1$Princeton Plasma Physics Laboratory, Princeton University, Princeton, NJ, USA}
\affiliation{$^2$Institute for Fusion Studies, University of Texas at Austin, Austin, Texas, USA}
\affiliation{$^3$ExoFusion, Austin, Texas, USA}
\affiliation{$^4$Department of Applied Physics and Applied Mathematics, Columbia University, New York, NY, USA}
\affiliation{$^5$York Plasma Institute, Department of Physics, University of York, Heslington, York, United Kingdom}
\begin{abstract}
H-mode operation of tokamak fusion plasmas free of dangerous Type 1 edge-localized-modes (ELMs) requires a non-ELM mechanism for saturating the edge pedestal growth. One possible mechanism is turbulent transport. We introduce a transport threshold model to find pedestal width-height scalings for turbulent transport-limited pedestals. The model is applied to electron heat transport resulting from electron-temperature-gradient (ETG) turbulence. The width-height scalings are highly sensitive to the relative contribution of density and temperature to the pedestal pressure. Pressure that builds up mainly through temperature is more likely to be transport-limited, and hence ELM-free. A relative radial inward shift of the temperature to density pedestal location is also more likely to transport-limit the pedestal. A second constraint such as flow shear is required to saturate pedestal growth. We also calculate width-height transport scalings resulting from particle and heat transport arising from ETG and kinetic-ballooning-mode turbulence. Comparisons are performed for ELMy and ELM-free experiments in MAST-U, NSTX, and DIII-D. This is a first step towards a pedestal width-height scaling for transport-limited ELM-free pedestals.
\end{abstract}

\maketitle

\setlength{\parskip}{0mm}
\setlength{\textfloatsep}{5pt}
\setlength{\belowdisplayskip}{6pt} \setlength{\belowdisplayshortskip}{6pt}
\setlength{\abovedisplayskip}{6pt} \setlength{\abovedisplayshortskip}{6pt}

\section{Introduction}

Edge-localized-modes (ELMs) are periodic eruptions of heat and particles from tokamak fusion plasmas into the scrape-off layer and divertor areas \cite{Zohm1996,Snyder2002,Kirk2004,Leonard2014}. ELMs are usually present in H-mode, a high-confinement regime for tokamaks \cite{Wagner1982,Groebner_2023}. Type 1 ELMs -- the most dangerous type of ELM -- while tolerable on current experiments, are not power exhaust-compatible in future tokamak power plants and must be avoided \cite{Maingi_2014,Hughes2020,Viezzer2023}. Given that the cost of a fusion power plant is predicted to be highly sensitive to confinement \cite{Wade2021}, H-mode is often assumed the default operating scenario tokamak power plant designs due to the benefits of higher confinement \cite{Federici2019b,Creely2020,Muldrew2024} (there are also notable benefits to L-mode operation \cite{Najmabadi_2006,Wilson2024,Rutherford_2024}). Given the confinement enhancement, operating in H-mode without large ELMs is a very high research priority.

There are promising candidates for ELM-free or small-ELM operation, each with their challenges, and no single one has yet emerged as a clear frontrunner \cite{Viezzer2023}. The quasi-continuous exhaust regime is a recently discovered small-ELM H-mode on ASDEX-U, characterized by ballooning instability at the separatrix \cite{Faitsch2021,Harrer_2022,Radovanovic_2022,Faitsch2023,Dunne2024}. Other small-ELM regimes include grassy ELMs \cite{Kamada_2000,Saibene_2005} and Type II ELMs \cite{Suttrop2000,Stober_2001,Sartori_2004}. Enhanced-pedestal H-mode was discovered on NSTX following lithium evaporation on plasma facing components, leading to ELM-free regimes with very wide pedestals \cite{Maingi2009,Maingi2010,Canik2013,Gerhardt_2014}. Enhanced D-alpha H-mode is ELM-free and is thought to be transport-limited by a quasi-coherent mode \cite{Hubbard2001,LaBombard_2014,Gil_2020,Macwan2024}. I-mode is an ELM-free regime characterized by L-mode-like particle confinement and H-mode-like temperature confinement \cite{Greenwald1998,Ryter1998,Hubbard_2017}. Quiescent H-mode (QH-mode) is an ELM-free regime where particle transport is believed to be sustained by an edge harmonic oscillation \cite{Burrell2001, Sakamoto_2004, Suttrop_2005, Garofalo_2011}. A variant of QH-mode, however, is hypothesized to be turbulence-limited \cite{Ernst2024}. Similarly, the related wide-pedestal QH-mode is thought to be governed by turbulence-limited mechanisms \cite{Burrell2016, Chen2017b, Wilks2021b, Houshmandyar2022}. There are also regimes with active ELM control using resonant magnetic perturbations \cite{Kirk_2010}, vertical kicks \cite{Degeling_2003}, ELM pacing \cite{Evans_2004}, and supersonic molecular beam injection \cite{LianghuaYao2001}. Negative triangularity \cite{Austin2019,Nelson2023,Guizzo2024,Wilson2024,Thome2024,Paz-Soldan_2024} has recently emerged as an attractive naturally ELM-free regime, where the plasma edge profiles are thought to be limited by ballooning modes \cite{Nelson2022,Nelson2024b}. L-mode is an important ELM-free regime, typically -- but not always -- characterized by lower core power density \cite{Jardin2000}.

For naturally ELM-free regimes, non-ELM mechanisms are required to limit and saturate the edge pedestal growth such as the quasi-coherent mode and edge harmonic oscillation mentioned above. In this work, we study how turbulent transport can constrain pedestal growth and give naturally ELM-free regimes.

We find scaling expressions for pedestal width and height in regimes where pedestal profiles are transport-limited rather than limited by MHD instabilities such as kinetic-ballooning-modes (KBMs) \cite{Tang1980,Aleynikova2018} and peeling-ballooning-modes (PBMs) \cite{Lortz1975,Connor1998,Wilson1999}. In this work, we focus on turbulent heat and particle transport resulting from electron-temperature-gradient (ETG) \cite{Lee1987,Horton1988,Dorland2000,Jenko2000,Zielinski2017,Adkins2022,Li_2024,Nasu_2024,Clauser2024,Ren2024} and KBM turbulence \cite{Pueschel2008,Maeyama2014,ishizawa2019,Mckinney2021,Kumar2021}. Because there are other transport mechanisms, the ETG and KBM-limited scalings are an approximate upper bound on the pressure in transport-limited pedestals, assuming the inward particle pinch is weak  \cite{Ware1970,Wagner_1993,Hoang2003,Weisen_2004}.

We show that while transport-limited pedestals can be more robustly ELM-free, another constraint such as $\mathbf{ E} \times \mathbf{ B}$ flow shear \cite{Hahm1995,McDermott2009,Barnes2011b} is required to provide a benign pedestal growth saturation mechanism. A major challenge in tokamak power plant design is core-edge integration of the plasma that includes a self-consistent ELM-free / small-ELM pedestal \cite{Snyder2012,Meneghini2016,Meyer2017,Meneghini2021,Viezzer2023,Snyder2024,Welsh2025}. This task would be simplified dramatically with simple expressions for ELM-free pedestal width and height. This work is a first step towards a reduced model for transport-limited ELM-free / small-ELM pedestals that ultimately can be used in integrated modeling for power plant design \cite{Romanelli2014,Meneghini2015,Imbeaux_2015,Artaud_2018,Staebler_2022,Badalassi2023,Lyons2023,Anand2023}.

\begin{figure}[tb!]
    \centering
    \begin{subfigure}[t]{0.39\textwidth}
    \centering
    \includegraphics[width=1.0\textwidth]{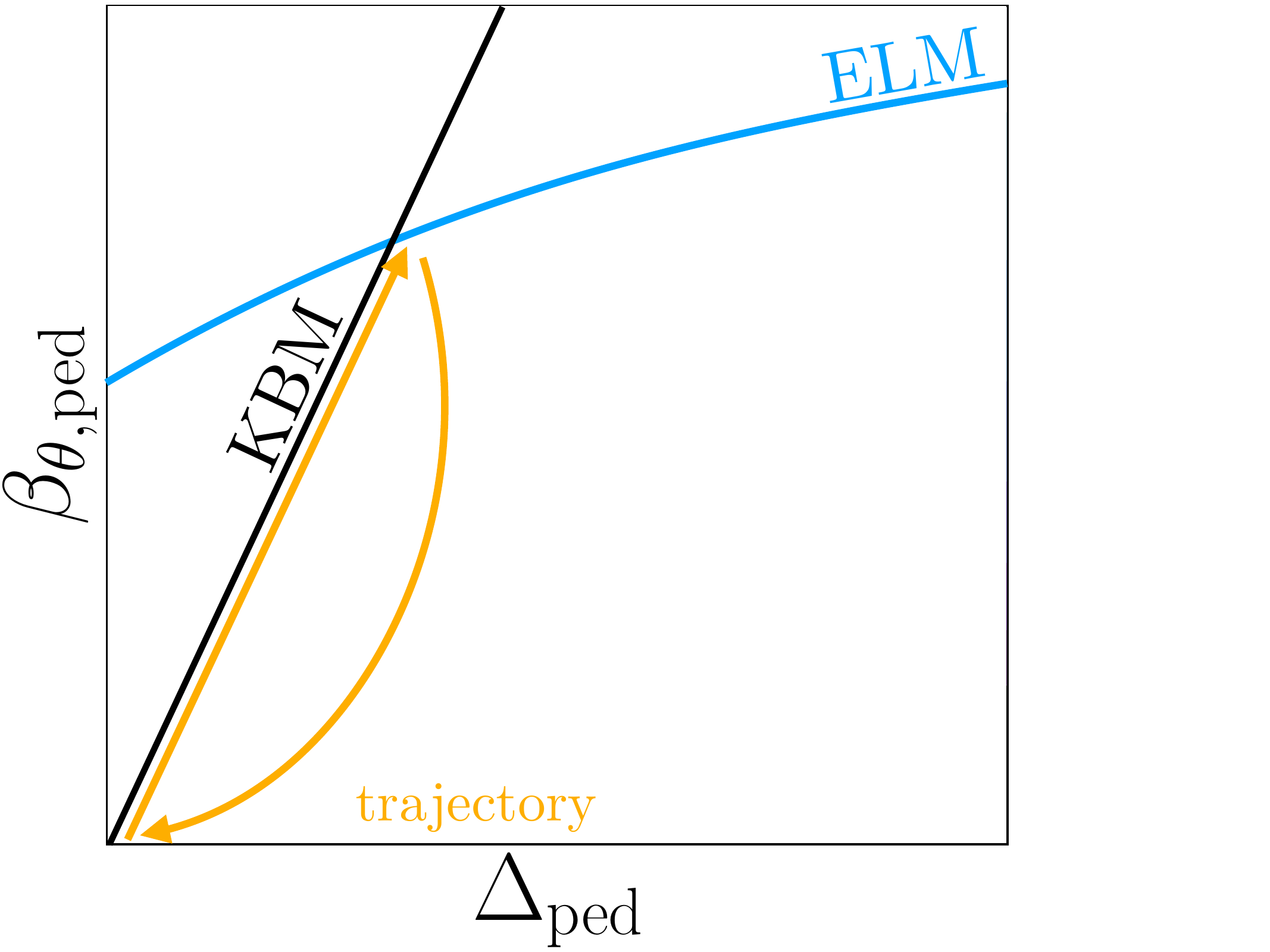}
    \caption{Trajectory for ELM and KBM-limited pedestal.}
    \end{subfigure}
    \begin{subfigure}[t]{0.39\textwidth}
    \centering
    \includegraphics[width=1.0\textwidth]{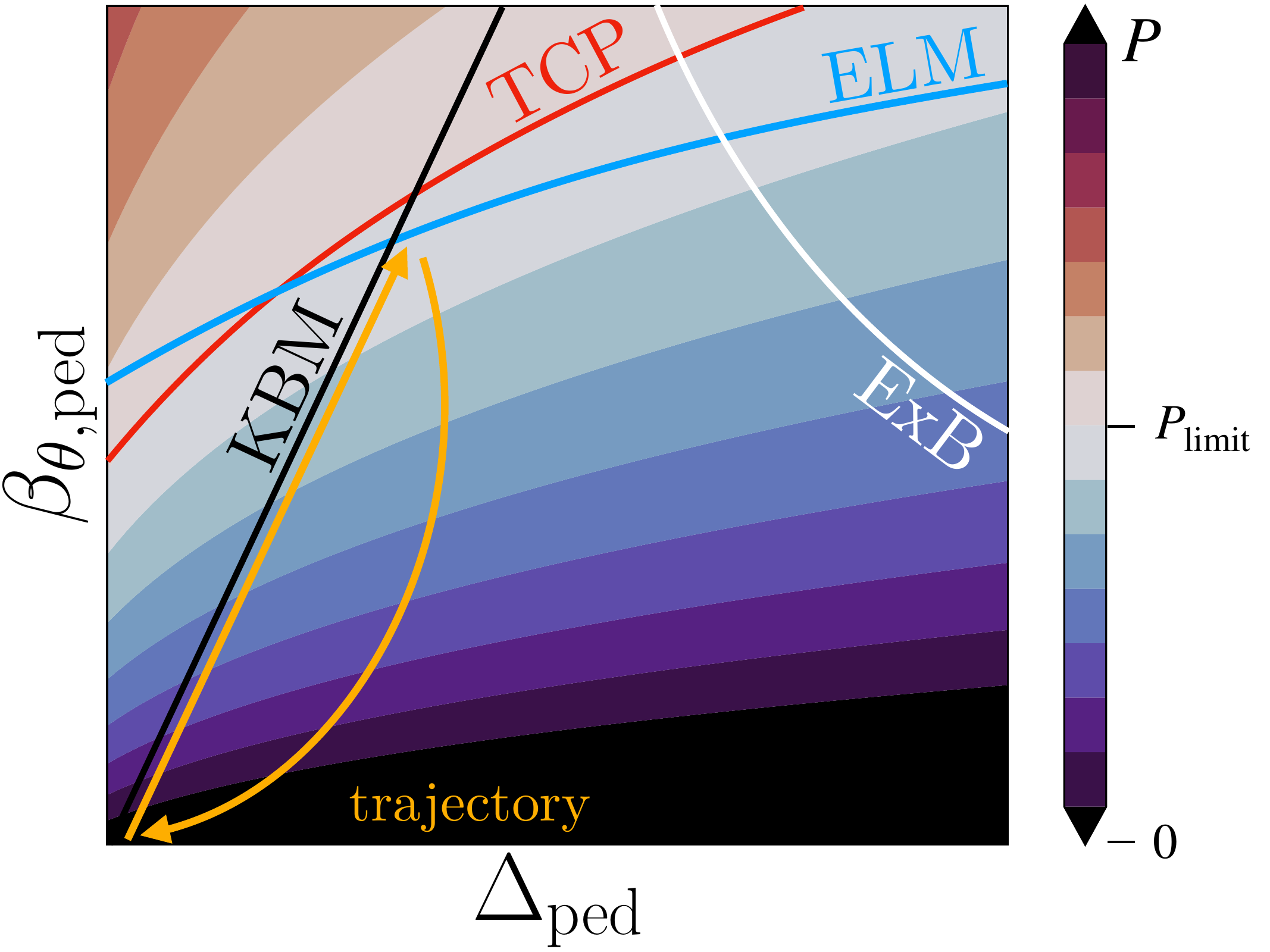}
    \caption{Trajectory for ELM and KBM-limited pedestal with transport.}
    \end{subfigure}
    \begin{subfigure}[t]{0.39\textwidth}
    \centering
    \includegraphics[width=1.0\textwidth]{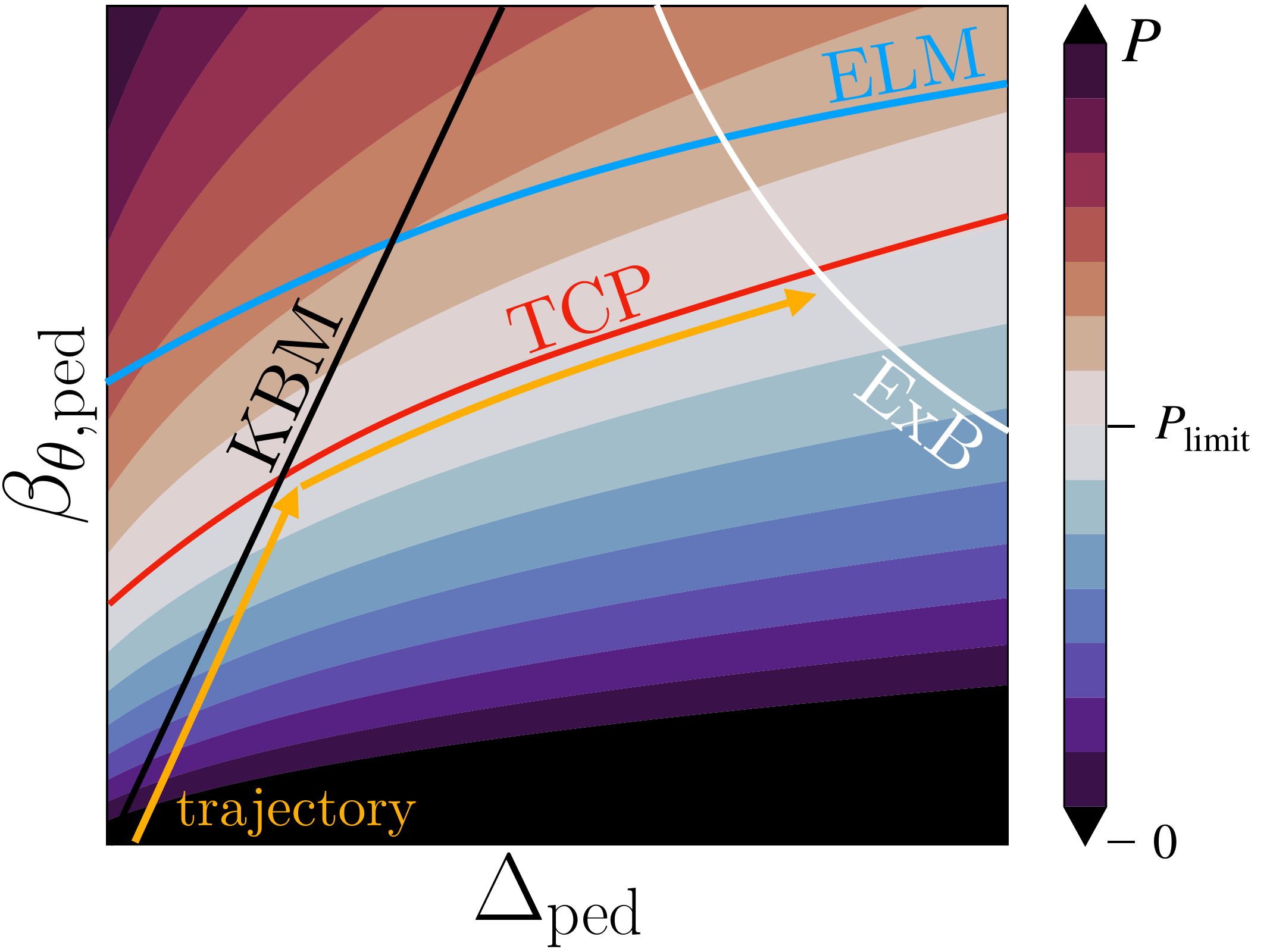}
    \caption{ELM-free trajectory for TCP and ExB-limited pedestal with transport.}
    \end{subfigure}
     ~
    \caption{Heuristic pedestal trajectories in width-height ($\Delta_{\mathrm{ped} }$,$\beta_{\theta, \mathrm{ped} }$) space for a type-I ELMy H-mode and a transport-limited H-mode. (a) standard picture of an ELM limiting highest pressure. (b) same trajectory as (a) but with Transport Critical Pedestal (TCP) and ExB flow shear constraint. (c) pedestal trajectory for a TCP and ExB flow shear-limited pedestal. $P$ is the power due to turbulence and $P_{\mathrm{limit}}$ is the limiting source power through the pedestal.}
    \label{fig:cartoon}
\end{figure}

Pedestal scalings for Type 1 ELMy H-modes can be predicted with EPED \cite{Wilson2004,Snyder2009}, a pedestal model that posits an ELMy pedestal's maximum achievable height is ultimately limited by an ELM that is predicted by PBM stability. In the EPED model, during the inter-ELM cycle before the pedestal pressure and current are sufficiently high to trigger an ELM, the pedestal's trajectory is KBM-limited \cite{Snyder2009}. The KBM-limited and post-ELM trajectories in pedestal width $\Delta_{\mathrm{ped} }$ and height $\beta_{\theta, \mathrm{ped} }$ space are described by the orange arrows in \Cref{fig:cartoon}(a). However, because Type 1 ELMy H-modes are not viable for tokamak power plants, a reduced pedestal model is needed for ELM-free H-modes. This paper shows how transport mechanisms in addition to KBM could steer the pedestal's trajectory \cite{Diallo_2021}, and in some cases, be used to avoid ELMs. By intentionally keeping the pedestal pressure far below the ELM-limit, transport-limited pedestals would be ELM-free. Similar to the arguments presented in \cite{Kotschenreuther_2024a,Kotschenreuther2024}, we also show how steep density gradients might enable pedestals with lower heating power than those with steep temperature gradients. However, care is required to ensure that such pedestals remain ELM-free.

While previous work has shown that KBM can limit pedestal width and height \cite{Snyder2009,Parisi_2024}, this may not hold for all pedestals, even Type 1 ELMy pedestals \cite{Dickinson2012,Saarelma2013,Diallo2015,Laggner2016,Maggi2017}. There are important differences between KBM and other gyrokinetic instabilities. First, while KBM is largely destabilized by the pressure gradient, other gyrokinetic instabilities are highly sensitive to the relative density and temperature gradients. Second, heat and particle transport arising from KBM is usually modeled as extremely stiff -- that is the linear critical gradient is the steepest gradient allowed by KBM. In contrast, in the absence of KBM, profile gradients steepen above the linear critical gradient for a given gyrokinetic instability and nonlinearities can even change the nonlinear critical gradient \cite{Dimits2000}. This means that a linear critical gradient model cannot predict the profiles in the presence of turbulent transport -- perhaps with the exception of KBM. Therefore, any model of pedestal width and height limited by gyrokinetic turbulence requires expressions for the fluxes above the linear critical gradient.

There are many gyrokinetic instabilities \cite{Krall1965,Drake1980,Guzdar1983,Dorland2000,Xiao2009,Roach2009,Garbet_2010,Kotschenreuther2019}. For simplicity, this work focuses on slab ETG -- which is destabilized by electron temperature gradients and stabilized by density gradients \cite{Lee1987,Dorland2000,Jenko2000,Adkins2022} -- and KBM. Using a recent model for the electron heat flux \cite{Hatch2024}, we calculate width-height scalings and estimate proximity to transport-limited pedestals for MAST-U, NSTX, and DIII-D discharges. %

The layout of this paper is as follows. We introduce the transport threshold model in \Cref{sec:TCP}. In \Cref{eq:ETGmodelintro}, we introduce an ETG model for parameterized pedestal profiles and calculate width-height scalings. We apply this model to MAST-U, NSTX, and DIII-D experiments in \Cref{sec:hatchmodel}. Prospects for ELM-free operation are discussed in \Cref{sec:ELMfree}. We introduce a combined ETG and KBM heat and particle transport model in \Cref{sec:combined_particle_heat_model} and calculate the resulting width-height scalings. Transport model extensions are described in \Cref{sec:modelextensions}. We summarize in \Cref{eq:discussion}. In \Cref{app:powerbalance,app:particlebalance}, we give examples of electron power and particle balance from an NSTX and DIII-D discharge.

\section{The Transport Critical Pedestal Constraint} \label{sec:TCP}

In this section, we describe how to obtain the width-height scaling for transport-limited pedestals using a transport threshold model. We call this scaling the Transport Critical Pedestal (TCP).

With accurate sources and flux models, one could find the time evolution of the density, momentum, and temperature pedestal profiles, giving the evolution of pedestal width and height. In ELM-free regimes, the mechanism limiting pedestal growth — transport limiting gradients — would be enforced by the profile gradients building up sufficiently such that the fluxes equal the sources. However, modeling this system is very challenging and beyond the scope of this work.

We instead use a lower fidelity model that aims to capture essential features of the higher fidelity model. We focus on electron energy transport. The electron energy transport equation for a heat transport-limited pedestal is
\begin{equation}
\frac{3}{2} \frac{\partial n_e T_e}{\partial t} = \sum_k S_{e,k} - \nabla \cdot \mathbf{q}_e = 0,
\end{equation}
where $n_e$ and $T_e$ are electron density and temperature, $t$ is time, $S_{e,k}$ are electron sources, and $q_e$ is electron heat flux. In this transport-limited pedestal, the turbulent power $P$ is equal to the net source power $P_{\mathrm{limit} }$ at some flux surface in the pedestal,
\begin{equation}
P \equiv q_e S = P_{\mathrm{limit} } \equiv \int \sum_k S_{e,k} d V,
\label{eq:powerbalance}
\end{equation}
where $S = 4 \pi^2 a R$ for minor radius $a$ and major radius $R$, and $dV$ is a volume integral. Assuming that $\sum_k S_{e,k}$ does not vary significantly over the pedestal, we will evaluate the right-hand side of \Cref{eq:powerbalance} at the separatrix.

We obtain transport width-height scalings with the following ansatz: once $P = P_{\mathrm{limit} }$ at any radial location in the pedestal, the pedestal height can no longer increase (at fixed pedestal width). For this transport-limited pedestal, we expect that if the pedestal widens further, it will increase in height until it hits a transport limit again. Therefore, the TCP is the set of $\Delta_{\mathrm{ped} },\beta_{\theta, \mathrm{ped} }$ values satisfying
\begin{equation}
P (\Delta_{\mathrm{ped} },\beta_{\theta, \mathrm{ped} }, \ldots) = P_{\mathrm{limit} }.
\label{eq:PCPcondition}
\end{equation}
Our model assumes that the profiles always satisfy a parameterized form (\Cref{eq:1,eq:2}) so that $\Delta_{\mathrm{ped} }$ and $\beta_{\theta, \mathrm{ped} }$ are always well-defined.

The arguments here might inform future experiments that manipulate transport to steer the pedestal trajectory far away from ELMs. In the context of recent insights \cite{Kotschenreuther_2024a} on transport-limited pedestals, this approach may validate ideas of how to achieve high pedestal pressure with low plasma heating power. While we focus on ETG instability, future work can include other instabilities and transport mechanisms, discussed in \Cref{sec:modelextensions}.

In \Cref{fig:cartoon}(b) and (c), we sketch trajectories of (b) an ELM and KBM-limited pedestal and (c) a TCP and $\mathbf{ E} \times \mathbf{ B}$ flow shear-limited pedestal. \Cref{fig:cartoon}(b) is the same as \Cref{fig:cartoon}(a), except with the transport power $P$ indicated with contours. For $P < P_{\mathrm{limit} }$, the temperature profile can increase because the sources exceed the transport. However, for $P = P_{\mathrm{limit} }$, the temperature can no longer increase and the pedestal is transport-limited. In (b), $P < P_{\mathrm{limit} }$ for all regions below the KBM and ELM limit, and therefore the pedestal is not transport-limited because the KBM or ELM limits are hit first. However, in \Cref{fig:cartoon} (c), the ETG power has a higher stiffness. Therefore, while at lower widths the pedestal is KBM-limited, at a certain point the ETG transport increases sufficiently to change the trajectory to being transport-limited, following the TCP.

An additional mechanism is required to saturate an ELM-free pedestal. Transport and KBM, combined or individually, are insufficient to provide an ELM-free pedestal. In \Cref{fig:cartoon}(b) and (c), we also sketch a constraint given by $\mathbf{ E} \times \mathbf{ B}$ flow shear \cite{Parisi_2024c}. If the $\mathbf{ E} \times \mathbf{ B}$ constraint intersects either the KBM or TCP before the ELM constraint, the pedestal could remain ELM-free. We assume that the intersection of the TCP and $\mathbf{ E} \times \mathbf{ B}$ constraints in \Cref{fig:cartoon} (c) is a stable point.

\section{ETG Transport Model} \label{eq:ETGmodelintro}

In this section, we calculate the pedestal width-height scalings resulting from a heat flux model for ETG transport. In this section, all of the work is algebraic. In subsequent sections, we compare with experiment. ETG instability \cite{Dorland2000,Jenko2000} is driven by electron temperature gradients. In the pedestal, ETG turbulence can be multiscale \cite{Maeyama2015,Howard2016,Hardman2019}, spanning both electron and ion gyroradius scales \cite{Parisi2020,Parisi2022,Belli_2023,Chapman2024}. It primarily transports heat through the electron channel, and the heat flux is typically much larger than the particle flux.

\subsection{Parameterized profiles}

We use a standard parameterization \cite{Snyder2009b,Parisi_2024} for the profiles,
\begin{equation}
\begin{aligned}
& n_e(\psi_N) = n_{e,\mathrm{core}} \mathrm{H} \left[  \psi_{\mathrm{ped,n_e}}  -  \psi_N \right] (1-\psi_N^{\alpha_{n_1}})^{\alpha_{n_2}} + \\
& A_n \bigg{[} n_{e0} \left( t_1 - \tanh \left(\frac{\psi_N-\psi_{\mathrm{mid,n_e}}}{ S_{\Delta} \Delta_{n_e} /2} \right) \right) \bigg{]} + n_{\mathrm{e,off}},
\label{eq:1}
\end{aligned}
\end{equation}
\begin{equation}
\begin{aligned}
& T_e(\psi_N) = T_{e,\mathrm{core}} \mathrm{H} \left[  \psi_{\mathrm{ped,T_e}} -  \psi_N \right] (1-\psi_N^{\alpha_{T_1}})^{\alpha_{T_2}} +  \\
& A_T \bigg{[} T_{e0} \left( t_1 - \tanh \left(\frac{\psi_N-\psi_{\mathrm{mid,T_e}}}{ S_{\Delta} \Delta_{T_e}/2} \right) \right) \bigg{]} + T_{\mathrm{e,off}},
\label{eq:2}
\end{aligned}
\end{equation}
where $H$ is a Heaviside function, $\psi_N$ is the poloidal flux normalized to 0 at the magnetic axis and 1 at the separatrix, $n_{e,\mathrm{core}}$, $T_{e,\mathrm{core}}$, $n_{e0}$, and $T_{e0}$ are constants, $\Delta_{n_e}$ and $\Delta_{T_e}$ are the pedestal electron density and temperature widths -- which often differ in experiments -- and ${\alpha_{\{n,T,J\},\{1,2\}}}$ are exponents. Quantities $A_{n}$, $A_{T}$, and $S_{\Delta}$ rescale the pedestal density, temperature, and width. The pedestal heights are
\begin{equation}
n_{e,\mathrm{ped}} = n (\psi_{\mathrm{ped,n_e}}), \; T_{e,\mathrm{ped}} = T(\psi_{\mathrm{ped,T_e}}),
\end{equation}
$n_{e,\mathrm{off}}$ and $T_{e,\mathrm{off}}$ are evaluated at $\psi_N = \psi_{\mathrm{ped,n_e}} + \Delta_{n_e}$ and $\psi_N = \psi_{\mathrm{ped,T_e}} + \Delta_{T_e}$ respectively, $t_1 = \tanh \left(1 \right)$, and $\psi_{\mathrm{ped},n_e} = \psi_{\mathrm{mid,n_e}} - \Delta_{n_e} / 2$. The pedestal width and pedestal top location are $\Delta_{\mathrm{ped}} = (\Delta_{n_e} + \Delta_{T_e})/2, \;\;\; \psi_{\mathrm{ped}} = \psi_{\mathrm{mid}} - \Delta_{\mathrm{ped}} / 2$, where $\psi_{\mathrm{mid}} =  (\psi_{\mathrm{mid},n_e}+ \psi_{\mathrm{mid},T_e})/2$ and $\beta_{\theta, \mathrm{ped}}$, $p_{ped}$ are  $\beta_{\theta, \mathrm{ped}} = 8 \pi p_{\mathrm{ped}} /\overline{B}_{\mathrm{pol}}^2, \;\; p_{\mathrm{ped}} = 2 p_e (\psi_N = \psi_{\mathrm{ped}} )$. Here, $\overline{B}_{\mathrm{pol}} =4\pi I_p /  l c $ with last-closed-flux-surface circumference $l$ where $I_p$ is the total plasma current and $c$ is the speed of light.

To describe increasing pedestal height with different density and temperature contributions, we use the pedestal pressure rescaling factor $S_p$ to rescale $p_{\mathrm{ped}}$,
\begin{equation}
S_p = \frac{p_{\mathrm{ped,new} }}{p_{\mathrm{ped,orig} }}  = S_T S_n.
\label{eq:Sp}
\end{equation}
Here, $S_T$ and $S_n$ are the temperature and density rescaling factors. We relate $S_T$ and $S_n$ using
\begin{equation}
S_T = \left( S_n \right)^b,
\label{eq:ST_Sn_scaling}
\end{equation}
where $b$ is a real number. For rescaling $p_{\mathrm{ped}}$ we use $S_T, S_n$ according to \Cref{eq:Sp} and $A_T, A_n$ (\Cref{eq:1,eq:2}), which are in general complicated functions, are calculated accordingly. Intuitively, as the pedestal pressure increases ($S_p > 1$), smaller $b$ corresponds to steeper density gradients. According to the arguments of density gradient turbulence stabilization \cite{Kotschenreuther_2024a,Kotschenreuther2024}, we expect pedestals with lower $b$ (most of the pressure pedestal increase comes from density) to support higher pedestals because less power is required to sustain such a pedestal. In this work we will not model the effects of density gradients on particle transport, but it is important to keep in mind that pedestals with steep density gradients are likely to hit a particle transport limit unless the particle source is very high.

\subsection{Transport Model}

\begin{figure}[!tb]
    \centering
    \begin{subfigure}[t]{0.4\textwidth}
    \centering
    \includegraphics[width=1.0\textwidth]{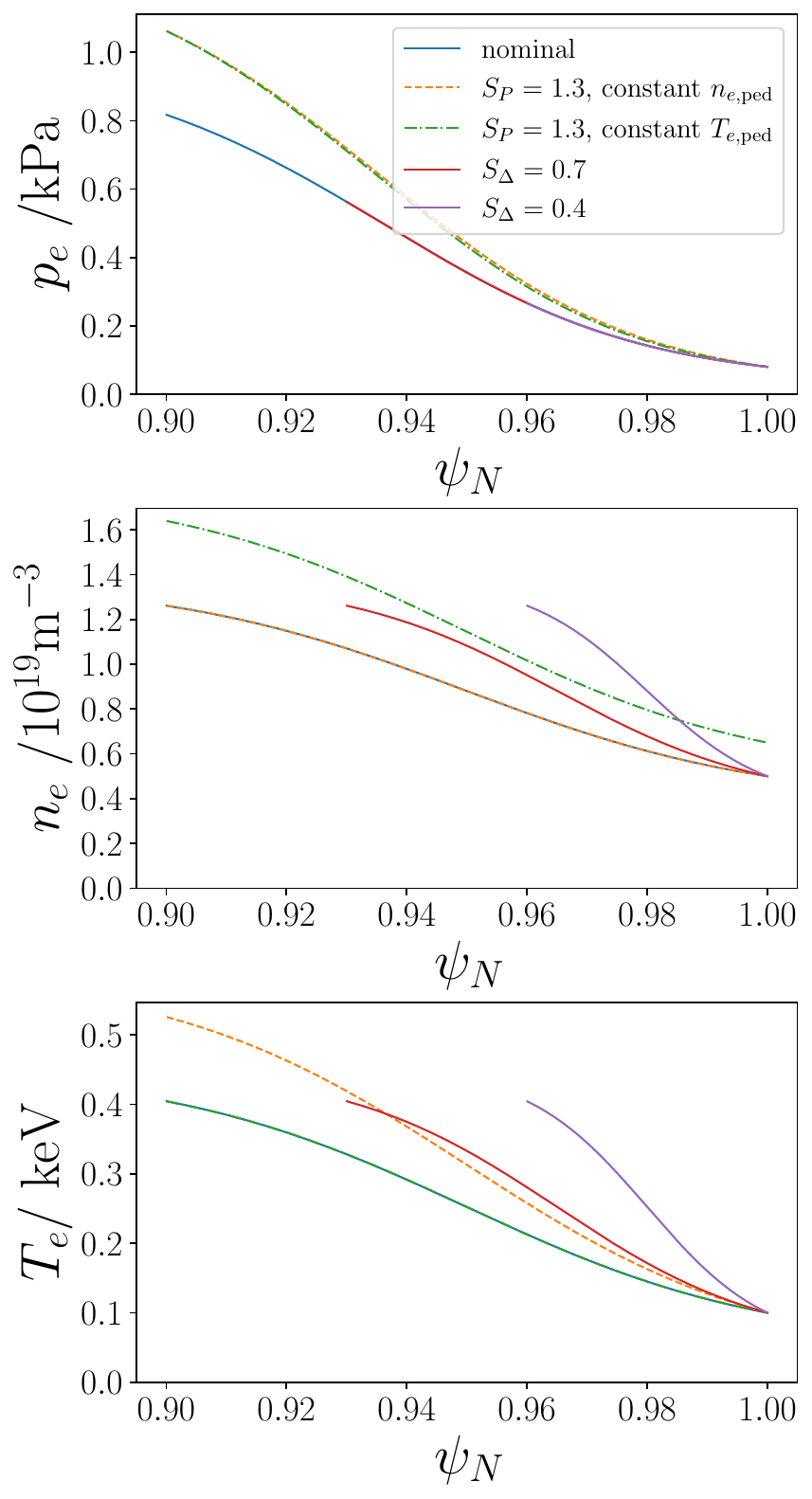}
    \caption{}
    \end{subfigure}
    \begin{subfigure}[t]{0.4\textwidth}
    \centering
    \includegraphics[width=1.0\textwidth]{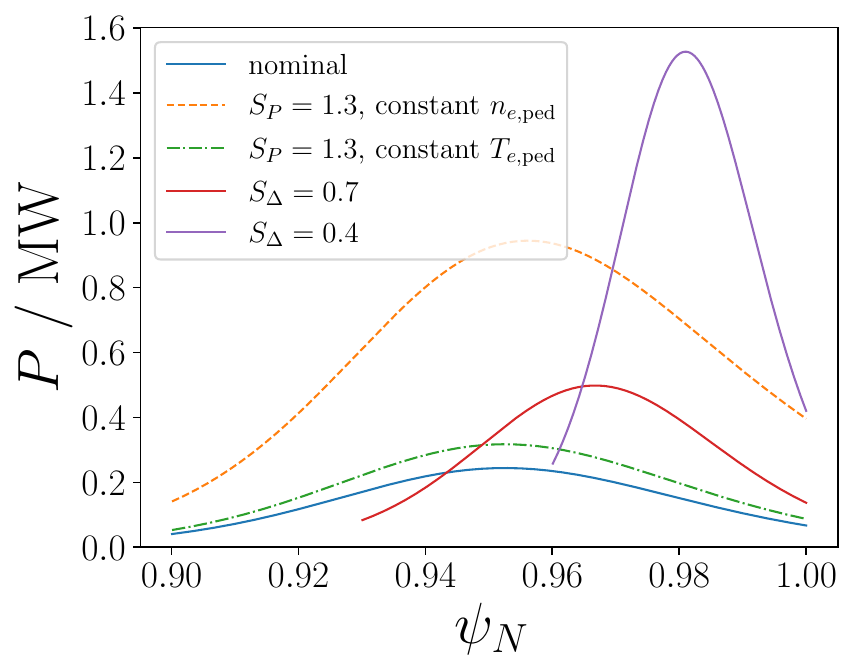}
    \caption{}
    \end{subfigure}
    \caption{(a) Electron pressure, density, and temperature profiles for the nominal and rescaled parameters in our analytic model, and (b) ETG power $P$.}
    \label{fig:profiles}
\end{figure}

We start with a heat flux model based on the ETG model in \cite{Hatch2024}, which describes slab ETG turbulence in the steep gradient region of the pedestal. Other works have shown the importance of toroidal ETG turbulence \cite{Parisi2022,Chapman2022,Chapman2024}. When applying the ETG flux model to an actual equilibrium in \Cref{sec:hatchmodel}, we use the exact same form as in \cite{Hatch2024}. The electron power per unit area is
\begin{equation}
q_e = A q_{gB} \omega_{T_e}^2 \eta_e^{1.54} \left( \eta_e - 1 \right) H \left(\eta_e - 1 \right).
\label{eq:heatflux2}
\end{equation}
Here, $A$ is an amplitude parameter, the gyroBohm heat flux normalization is $q_{gB} = n_e T_e c_s \rho_s^2 /a$ where $c_s = \sqrt{T_e/m_i}$, $\rho_s = \sqrt{T_e m_i / e^2 B^2}$, $e$ is the proton charge, $B$ is the magnetic field strength, and $a$ is the minor radius. The temperature-to-density gradient ratio is \cite{Jenko2000,Chapman2022,Guttenfelder2022,Hatch2024},
\begin{equation}
\eta_e = \omega_{T_e}/\omega_{n_e},
\end{equation}
where $\omega_{T_e} = - d \ln T_e / d \psi_N$, $\omega_{n_e} = - d \ln n_e / d \psi_N$. $\eta_{e,\mathrm{crit}}$ is the critical $\eta_{e}$ value required for instability.

We now calculate $P$ for model profiles. The nominal parameters we use are $A_T = 1.0$, $T_{e,\mathrm{off} } = 100$ eV, $T_{e,0} = 200$ eV, $A_n = 1.0$, $n_{e,\mathrm{off} } = 5 \times 10^{18}$ m${}^{-3}$, $n_{e,0} = 5 \times 10^{18}$ m${}^{-3}$, $\Delta_{\mathrm{ped} } = 0.03$, $a = 1.0$m, $B = 1.0$ T, $S = 10.0$ m${}^2$, $A = 1/60$, and $\eta_{e,\mathrm{crit}} = 1.0$. In \Cref{fig:profiles}(a), we plot the nominal electron pressure, density, and temperature pedestal profiles for these parameters. The resulting ETG power profiles are shown in \Cref{fig:profiles}(b). Of particular note, as the pedestal width is decreased at constant height, the electron power increases nonlinearly. This is shown by the $S_{\Delta} = 0.7$ and $S_{\Delta} = 0.4$ cases. This nonlinear scaling of power with width is an indication that there will be a nonlinear pedestal width-height scaling relation for transport-limited pedestals.

\begin{figure}[!tb]
    \centering
    \begin{subfigure}[t]{0.4\textwidth}
    \centering
    \includegraphics[width=1.0\textwidth]{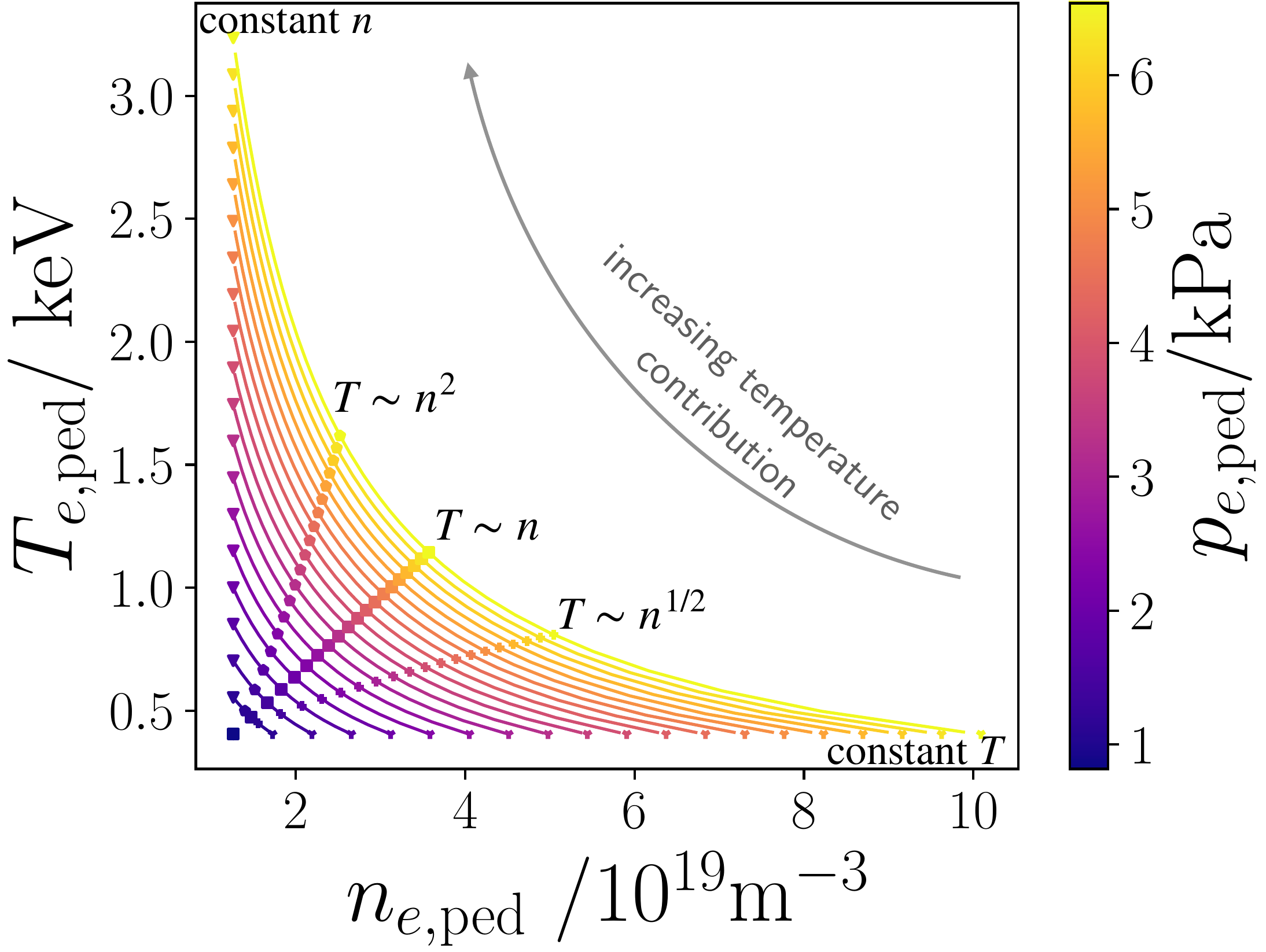}
    \caption{}
    \end{subfigure}
    \begin{subfigure}[t]{0.4\textwidth}
    \centering
    \includegraphics[width=1.0\textwidth]{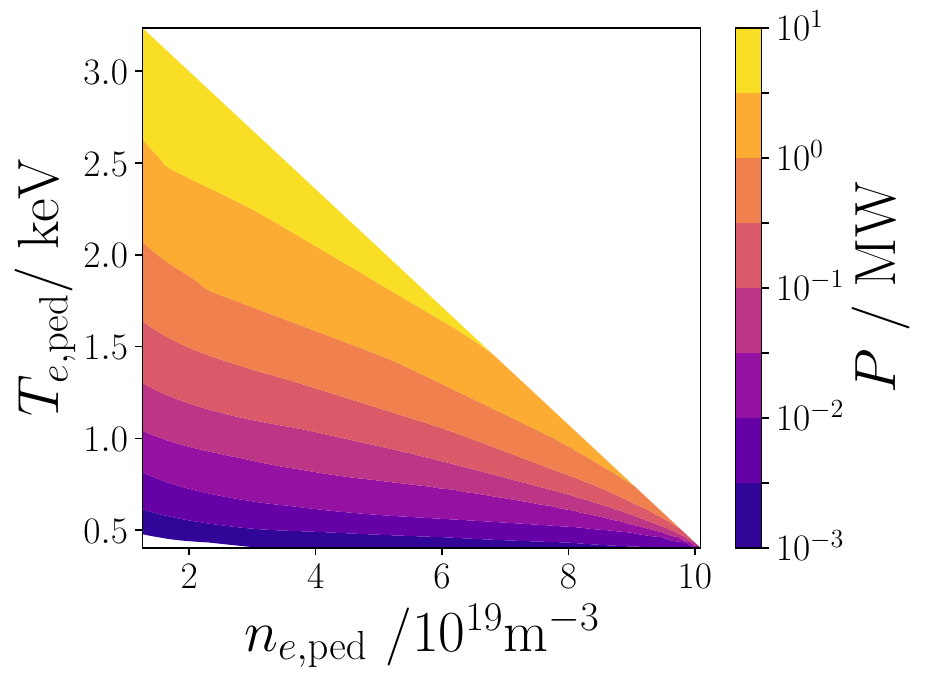}
    \caption{}
    \end{subfigure}
    \begin{subfigure}[t]{0.4\textwidth}
    \centering
    \includegraphics[width=1.0\textwidth]{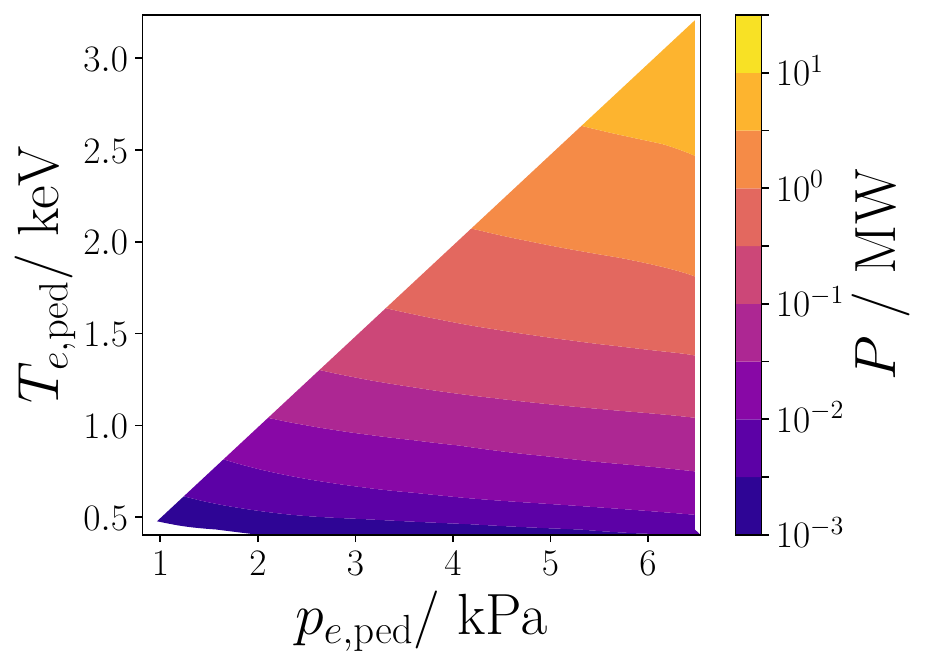}
    \caption{}
    \end{subfigure}
     ~
    \caption{Heat transport from ETG turbulence at fixed pedestal width: (a) contours of constant $p_{e,\mathrm{ped} }$ versus $n_{e,\mathrm{ped} }$ and $T_{e,\mathrm{ped} }$ with different height scalings according to \Cref{eq:ST_Sn_scaling}, (b) and (c): ETG power $P$ versus $n_{e,\mathrm{ped} }$, $T_{e,\mathrm{ped} }$, and $p_{e,\mathrm{ped} }$.}
    \label{fig:nped_Tped}
\end{figure}

Before calculating the width-height scalings, we note that there are an infinite number of ways to vary $n_{e,\mathrm{ped} }$ and $T_{e,\mathrm{ped} }$ at fixed pressure height $p_{e,\mathrm{ped} }$. This is shown in \Cref{fig:nped_Tped}(a) where the pedestal density and temperature heights are varied, and contours of constant $p_{e,\mathrm{ped} }$ have a fixed color. Different markers represent different relative contributions of density and temperature. In the analytic part of this work, we assume that
\begin{equation}
p_{\mathrm{ped}} \sim \beta_{\theta,\mathrm{ped}},
\end{equation}
which holds with the definition of $\beta_{\theta,\mathrm{ped}}$ if $I_p$ and the last-closed-flux-surface circumference stay constant across pedestal width and height.

\begin{figure}[!tb]
    \centering
    \begin{subfigure}[t]{0.4\textwidth}
    \centering
    \includegraphics[width=1.0\textwidth]{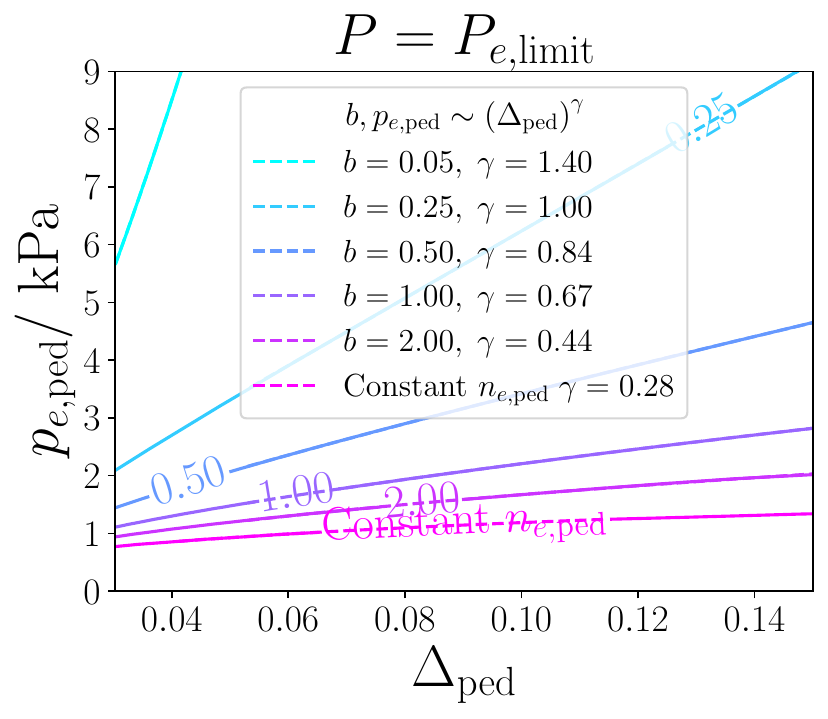}
    \caption{$n_{e,\mathrm{ped}}/n_{e,\mathrm{sep}}$ fixed.}
    \end{subfigure}
    \begin{subfigure}[t]{0.4\textwidth}
    \centering
    \includegraphics[width=1.0\textwidth]{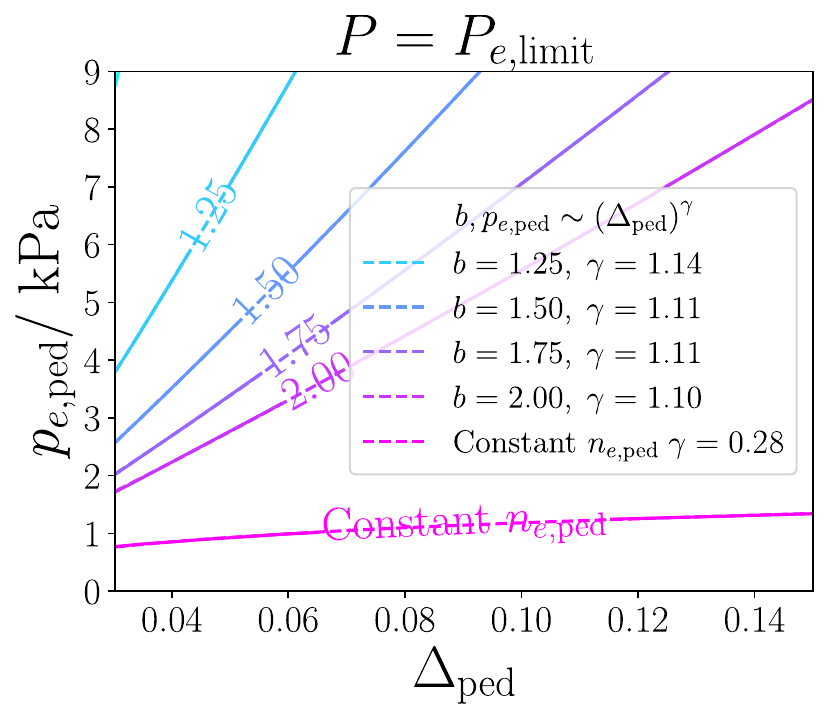}
    \caption{$n_{e,\mathrm{sep}}$ fixed. }
    \end{subfigure}
     ~
    \caption{Transport Critical Pedestal width-height scalings for a) $n_{e,\mathrm{ped}}/n_{e,\mathrm{sep}}$ fixed b) $n_{e,\mathrm{sep}}$ fixed. Scalings are labeled with $b$ value (see \Cref{eq:ST_Sn_scaling}) where higher $b$ results in larger density contributions to the total pedestal pressure.}
    \label{fig:width_height_scaling}
\end{figure}

The temperature and density profiles are scaled differently: when rescaling the temperature pedestal, we hold $T_{e,\mathrm{sep} }$ (which is $T_e$ evaluated at $\psi_N = 1.0$) constant. When rescaling the density pedestal, by default we hold $n_{e,\mathrm{ped} }/n_{e,\mathrm{sep} }$ constant. In \Cref{fig:profiles}(b), we show the effect on the radial profile of $P$ of rescaling the pedestal height with either constant $n_{e,\mathrm{ped} }$ or constant $T_{e,\mathrm{ped} }$. When rescaling density with $n_{e,\mathrm{ped} }/n_{e,\mathrm{sep} }$, the $\eta_e$ profile does not change, but $q_{gB}$ does linearly. Therefore, for $S_P = 1.3$, constant $T_{e,\mathrm{ped} }$ in \Cref{fig:profiles}(b), the peak value of $P$ only increases by a factor of $1.3$ because $q_{gB}$ increased by 1.3. However, if $p_{e,\mathrm{ped}}$ is increased with constant $n_{e,\mathrm{ped} }$, the peak power increases by much more. This is because not only does $q_{gB}$ increase, but $\omega_{Te}$ and $\eta_e$ also increase, increasing the heat flux as per \Cref{eq:heatflux2}, and befitting the behavior of electron-temperature-gradient instability.

We plot the peak value (across a radial profile) of $P$ in \Cref{fig:nped_Tped}(b), with the same axes of $T_{e,\mathrm{ped} }$ and $n_{e,\mathrm{ped} }$ as in \Cref{fig:nped_Tped}(a). The power increases most quickly when the pedestal pressure increases for increasing temperature at constant density. In \Cref{fig:nped_Tped}(c), we plot the same datapoints but replace $n_{e,\mathrm{ped} }$ on the y-axis with $p_{e,\mathrm{ped} }$. This demonstrates how $P$ might vary by orders of magnitude at constant $p_{e,\mathrm{ped} }$ depending on the relative density and temperature heights.

We now calculate the TCP for the analytic model, using the maximum available electron power $P_{e,\mathrm{limit}} = 2 \times 10^5 $W. In \Cref{fig:width_height_scaling}(a), we show the TCP for different relative contributions of density and temperature to the pedestal height buildup (corresponding to different values of $b$ in \Cref{eq:ST_Sn_scaling}), assuming that $n_{e, \mathrm{ped} }/n_{e, \mathrm{sep} }$ is fixed. The TCP scaling with the lowest pedestal gradients occurs for constant $n_{e, \mathrm{ped} }$. Rescaling the pedestal height with a relatively small density contribution ($b =2$) gives a TCP with nearly double the gradients of the constant $n_{e, \mathrm{ped} }$ case. Further decreasing $b$ gives significantly steeper pedestals according to the TCP. This is because decreasing $b$ causes density to increase much faster than temperature, so, as in \Cref{fig:nped_Tped}, the ETG power hits the $P_{e,\mathrm{limit} }$ at a higher $n_{e,\mathrm{ped} }$ and therefore higher $p_{e,\mathrm{ped} }$. Notably, \Cref{fig:width_height_scaling}(a) shows for $b = 2.00$, the pedestal height scales as $\Delta_{\mathrm{ped} } \sim \left( p_{e,\mathrm{ped} } \right)^2$. This means that small increases in pressure lead to larger increases in width, giving pedestals with lower average pressure gradient.

In \Cref{fig:width_height_scaling}(b), we show the TCP with fixed $n_{e, \mathrm{sep} }$, rather than fixed $n_{e, \mathrm{ped} }/n_{e, \mathrm{sep} }$. Because $\eta_e$ now decreases with increasing $n_{e, \mathrm{ped} }$, density gradient stabilization effects significantly increase the maximum achievable pedestal pressure. \Cref{fig:width_height_scaling}(b) reflects arguments \cite{Kotschenreuther_2024a,Boyle_2023,Maan2023,Berkery_2024} where low $n_{e, \mathrm{sep} }$ and high $T_{e, \mathrm{sep} }$ can achieve very high pedestal pressures with low heating power with the use of lithium, which often suppresses ELMs. This approach may not work for the quasi-continuous-exhaust regime where higher $n_{e, \mathrm{sep} }$ is typically desired \cite{Eich2024arxiv}.

\begin{figure}[!tb]
    \centering
    \begin{subfigure}[t]{0.4\textwidth}
    \centering
    \includegraphics[width=1.0\textwidth]{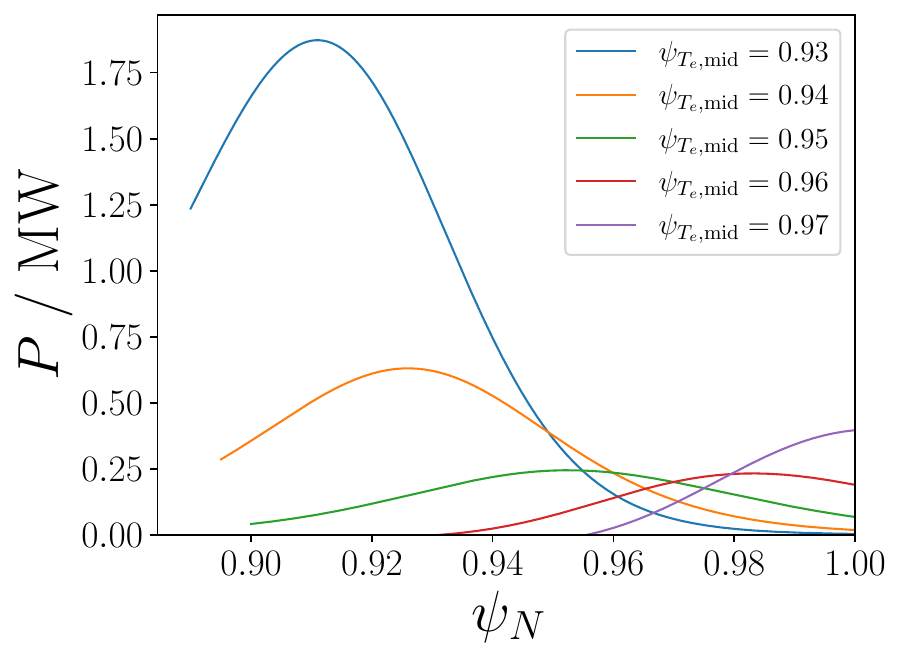}
    \end{subfigure}
    \caption{Radial profiles of $P$ for five pedestal profiles with fixed $ \Delta_{T_e,\mathrm{ped} } = \Delta_{n_e,\mathrm{ped} } = 0.1$, but with the temperature pedestal shifted relative to the density pedestal. Here, $\psi_{n_e,\mathrm{mid} } = 0.95$ is fixed.}
    \label{fig:heat_flux_examples_width_ratios}
\end{figure}

While the relative contribution of density and temperature to $p_{e,\mathrm{ped} }$ significantly affects the power, a relative shift of the temperature and density profiles also has a very large effect -- even when keeping $n_{e,\mathrm{ped} }$ and $T_{e,\mathrm{ped} }$ fixed. Gyrokinetic analysis of JET-ILW and JET-C discharges found the radial outward shift of density profiles in JET-ILW significantly increased the power required to sustain a pedestal \cite{Hatch2019}. In \Cref{fig:heat_flux_examples_width_ratios}, we show the effect on the radial ETG power profiles for a relatively shifted temperature pedestal. This is performed by varying $\psi_{T_e, \mathrm{mid} }$ between 0.93-0.97, while keeping all other parameters fixed, including $\psi_{n_e, \mathrm{mid} } = 0.95$. For example, the peak power in the mis-aligned $\psi_{T_e, \mathrm{mid} } = 0.93$ case is over seven times greater than the aligned $\psi_{T_e, \mathrm{mid} } = 0.95$ case. Note that $p_{e,\mathrm{ped} }$ is slightly higher for the radially inwards shifted temperature pedestal, but $p_{e,\mathrm{ped} }$ only varies by 10\% between all the shifted pedestals. The effect of radially shifted temperature pedestals will be important for the MAST-U discharges we study in the next section.

\begin{figure*}[!tb]
    \centering
    \begin{subfigure}[t]{0.47\textwidth}
    \centering
    \includegraphics[width=1.0\textwidth]{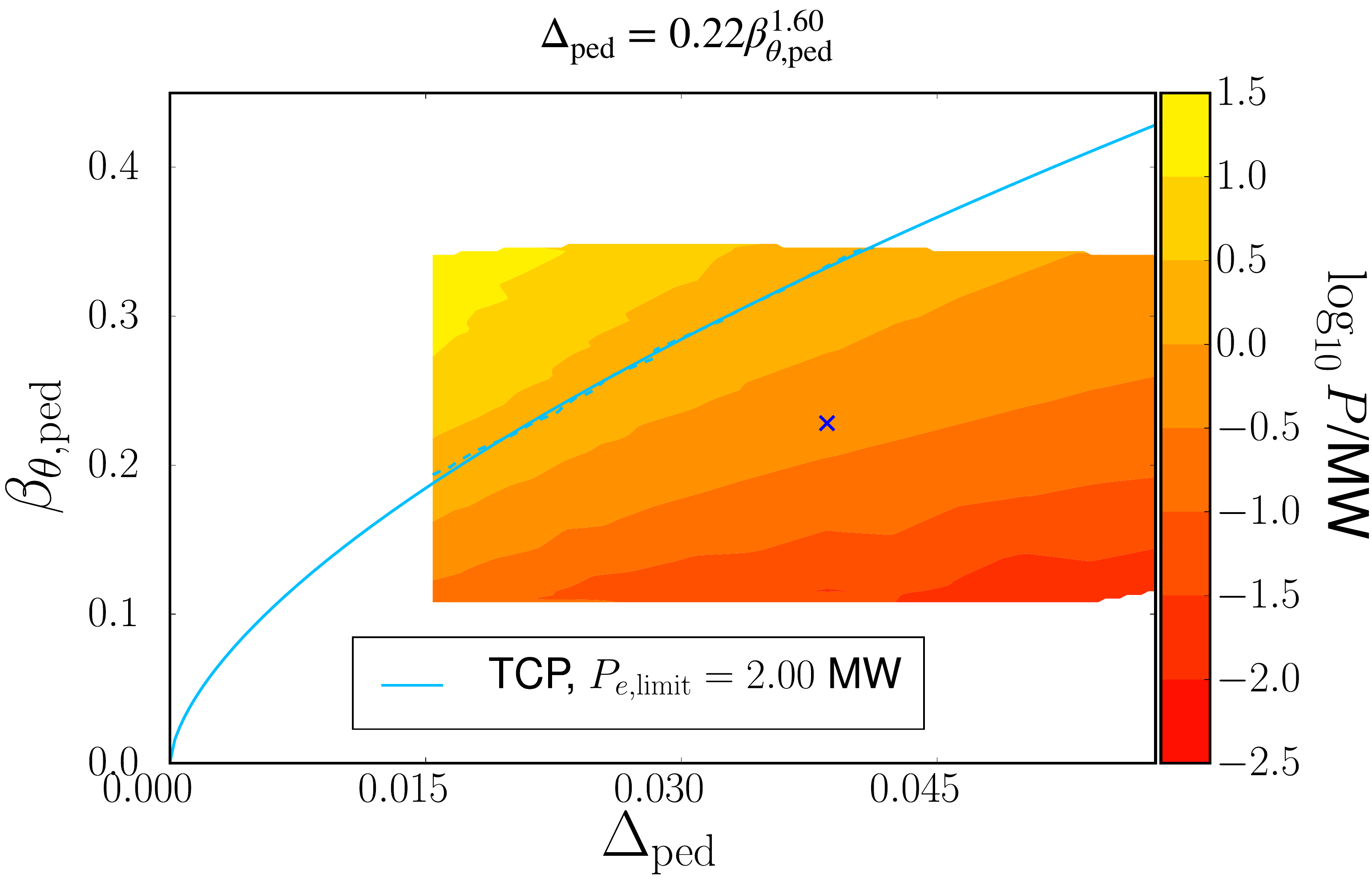}
    \caption{DIII-D 162940, constant $n_{e,\mathrm{ped}}$}
    \end{subfigure}
    \begin{subfigure}[t]{0.47\textwidth}
    \centering
    \includegraphics[width=1.0\textwidth]{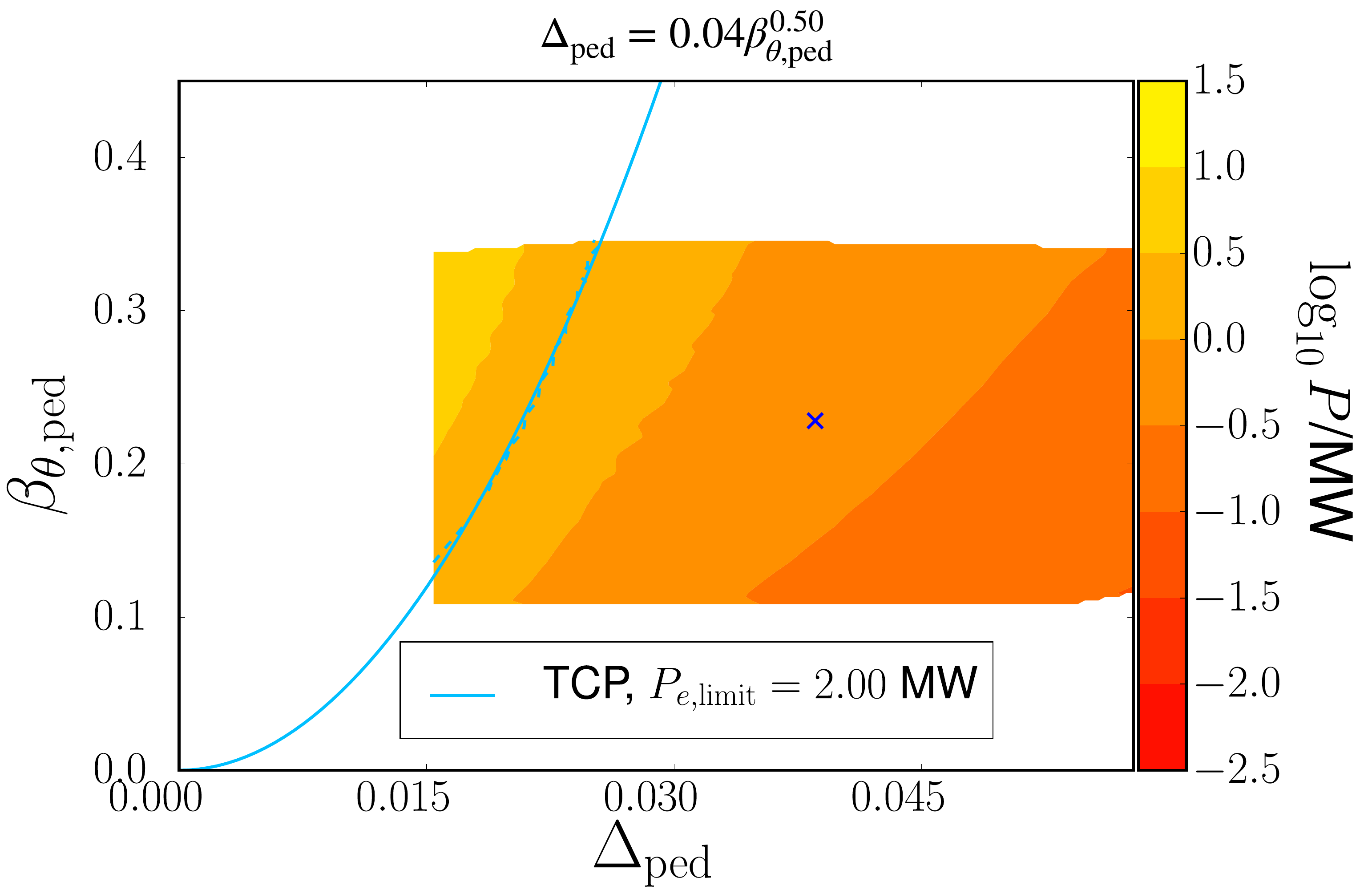}
    \caption{DIII-D 162940, constant $T_{e,\mathrm{ped}}$}
    \end{subfigure}
    \begin{subfigure}[t]{0.47\textwidth}
    \centering
    \includegraphics[width=1.0\textwidth]{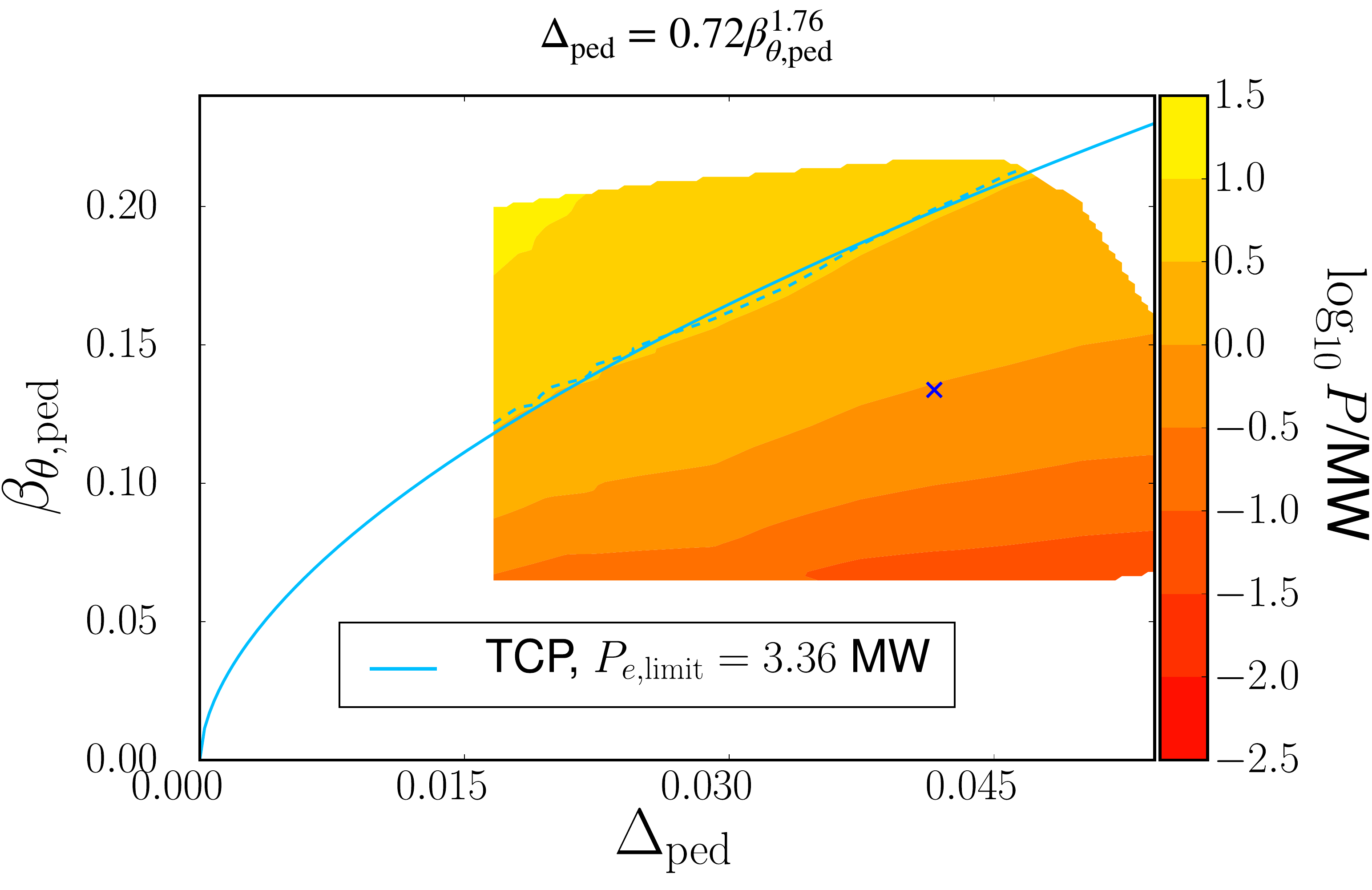}
    \caption{DIII-D 193843, constant $n_{e,\mathrm{ped}}$}
    \end{subfigure}
    \begin{subfigure}[t]{0.47\textwidth}
    \centering
    \includegraphics[width=1.0\textwidth]{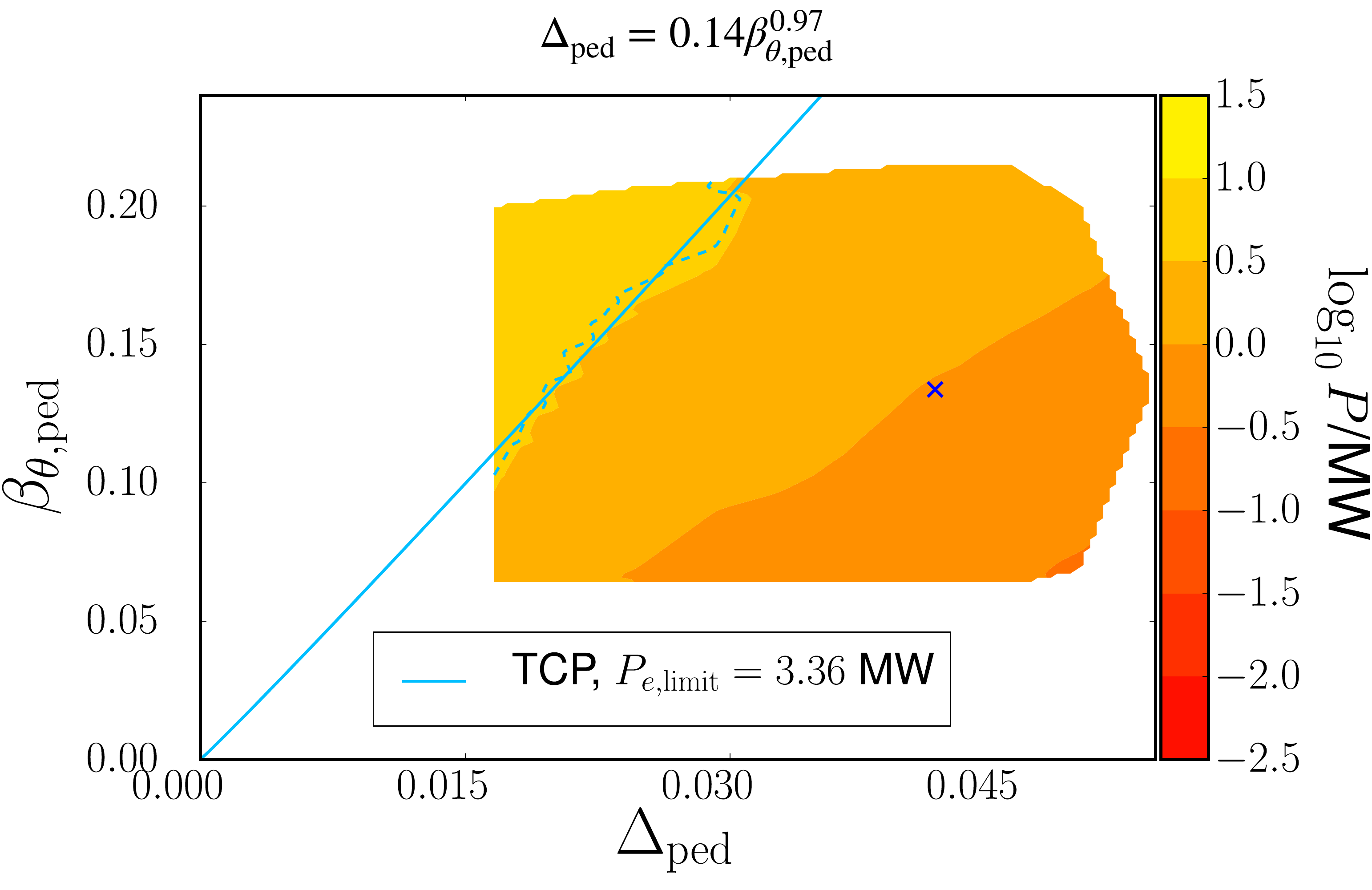}
    \caption{DIII-D 193843, constant $T_{e,\mathrm{ped}}$}
    \end{subfigure}
    \caption{TCP scalings and ETG power $P$ for DIII-D 162940 and 193843, with either constant $T_{e,\mathrm{ped}}$ or $n_{e,\mathrm{ped}}$. The cross marker is the experimental point.}
    \label{fig:width_height_DIIID}
\end{figure*}

\section{Experimental Comparison} \label{sec:hatchmodel}

We now perform the exercise above on experimental equilibria using a more realistic model for the electron heat flux \cite{Hatch2024},
\begin{equation}
q_e / q_{gB} = 0.1 \sqrt{\frac{m_e}{m_i}} \omega_{Te}^2 (\eta_e - 1) \frac{\eta_e^{1.54}}{ \tau^{0.5}} H \left( \eta_e - 1 \right),
\label{eq:qHatch23}
\end{equation}
where $\tau = Z_\mathrm{eff} T_e / T_i$. In order to calculate the TCP, we generate a self-consistent set of profiles and equilibria with varied pedestal width and height as in \cite{Parisi_2024,Parisi_2024b,Parisi_2024c}. We apply \Cref{eq:qHatch23} and the methodology in \Cref{sec:TCP} to find the TCP for DIII-D, MAST-U, and NSTX, experiments. For the ELMy DIII-D discharge, we also compare the TCP, KBM, and PBM constraints.

\subsection{DIII-D}

We first find TCP width-height scalings for two DIII-D discharges: 162940 at 2944ms, a well-benchmarked ELMy discharge \cite{Guttenfelder2021}, and negative triangularity L-mode ELM-free discharge 193843 at 3800ms \cite{Nelson2024b}. Despite 193843 being in L-mode, its edge temperature and density profiles are still well-approximated by the parameterizations in \Cref{eq:1,eq:2}, and therefore we proceed to find a TCP width-height scaling. For both DIII-D discharges, the TCP is calculated separately for constant $n_{e,\mathrm{ped} }$ and constant $T_{e,\mathrm{ped} }$.

The TCP for DIII-D 162940 with constant $n_{e,\mathrm{ped} }$, as before, is shown in \Cref{fig:width_height_DIIID}(a). We also show the TCP in \Cref{fig:width_height_DIIID}(b) with constant $T_{e,\mathrm{ped} }$. The corresponding TCP scalings are $\Delta_{\mathrm{ped} } = 0.25 \beta_{\theta, \mathrm{ped} }^{1.60}$ and $\Delta_{\mathrm{ped} } = 0.05 \beta_{\theta, \mathrm{ped} }^{0.48}$. We find $P$ is roughly 1/3 of the total electron power for DIII-D 162940. This is consistent with a previous analysis of the electron heat transport resulting from ETG \cite{Guttenfelder2021,Hatch2024}, which found roughly 1.5 - 2.5 MW of electron heat flux through the pedestal. Hence we estimate $P_{e,\mathrm{limit} } = 2.0 $MW. Since this is an ELMy discharge, we do not expect it to be close to being transport-limited, though. For constant $n_{e,\mathrm{ped} }$, the experimental pedestal height could increase by 50\% at fixed width until it hits the predicted ETG power limit.

The TCP for DIII-D 193843 is shown in \Cref{fig:width_height_DIIID}(c) and (d). For constant $n_{e,\mathrm{ped} }$, the TCP is $\Delta_{\mathrm{ped} } = 1.26 \left( \beta_{\theta, \mathrm{ped} } \right)^{1.81} $. For constant $T_{e,\mathrm{ped} }$, the TCP is $\Delta_{\mathrm{ped} } = 0.18 \left( \beta_{\theta, \mathrm{ped} } \right)^{0.83} $. The calculated ETG power is $P = 1.0$ MW. We have estimated the source power is $P_{e,\mathrm{limit} } = 3.36 $MW for this discharge. However, the equilibrium point is still relatively far from the TCP, even for the constant $T_{e,\mathrm{ped} }$ case in \Cref{fig:width_height_DIIID}(c). This indicates that additional transport mechanisms are required to explain the equilibrium point, or that another mechanism such as high-n ballooning modes \cite{Snyder2009,Nelson2024b,Parisi_2024} constrain the pedestal profiles.

\begin{figure}[!bt]
    \centering
    \begin{subfigure}[t]{0.43\textwidth}
    \centering
    \includegraphics[width=1.0\textwidth]{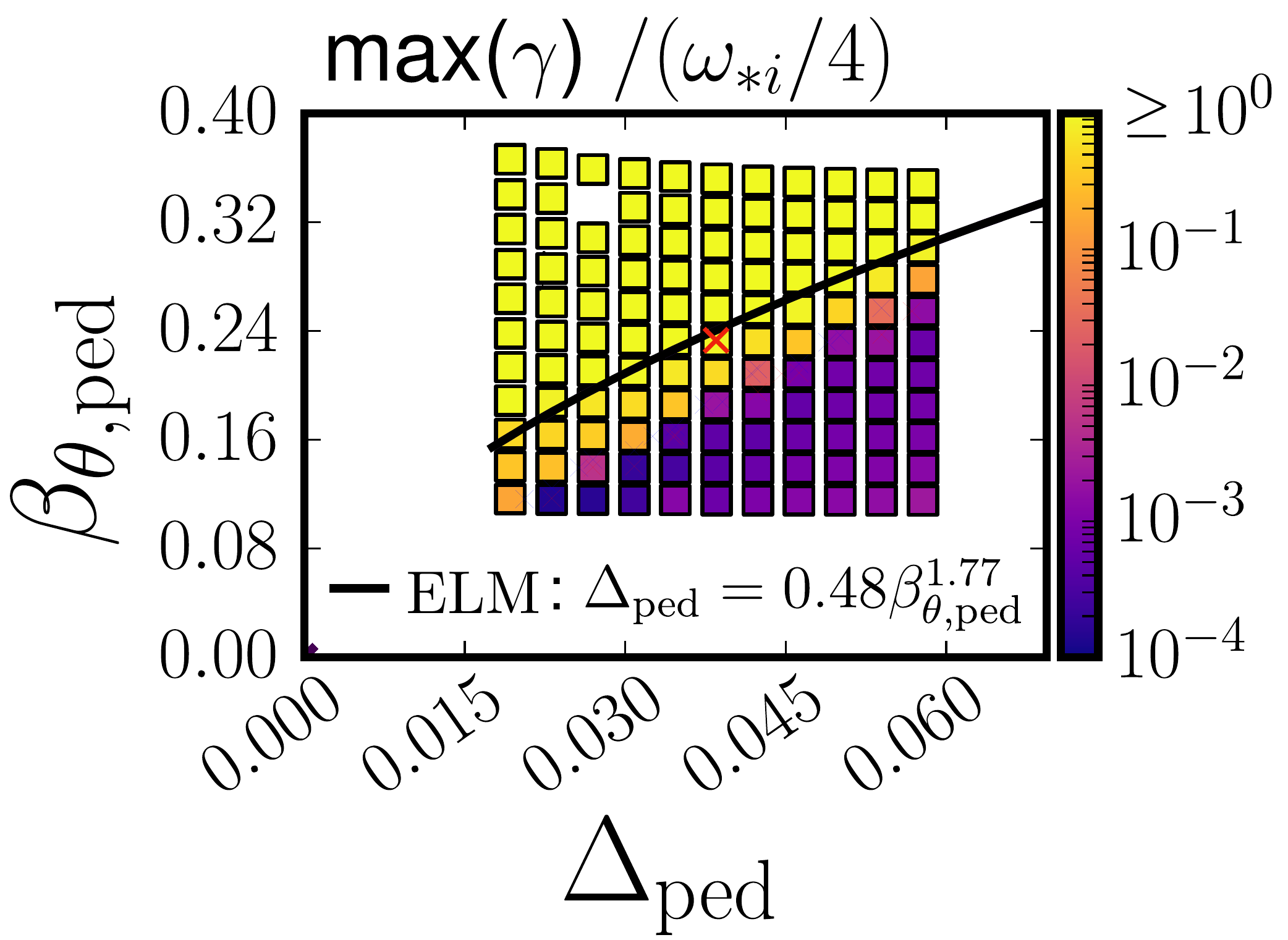}
    \caption{}
    \end{subfigure}
    \begin{subfigure}[t]{0.35\textwidth}
    \centering
    \includegraphics[width=1.0\textwidth]{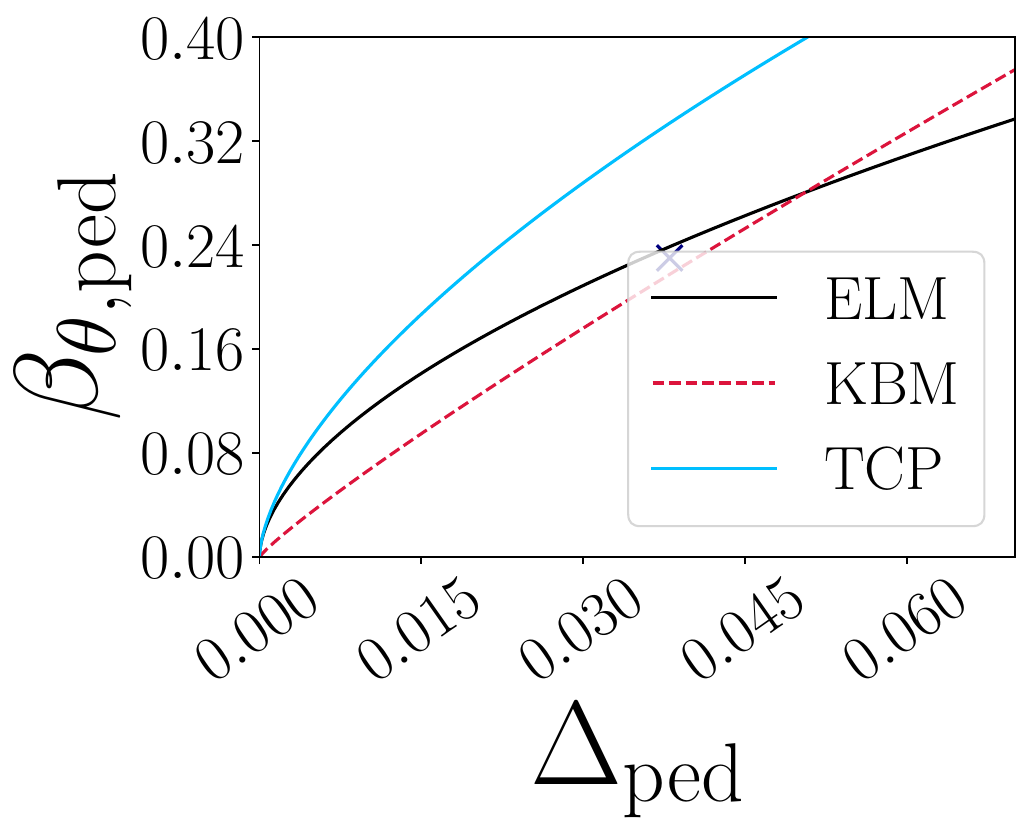}
    \caption{}
    \end{subfigure}
    \caption{(a) Growth rate of fastest growing ideal peeling-ballooning-mode for DIII-D 162940 constant $n_{e,\mathrm{ped}}$. The cross marker is the experimental point. The mode is unstable for $\gamma > \omega_{*i} /4$. (b) combined TCP, KBM, and ELM constraints.}
    \label{fig:width_height_ELM_D3D_162940}
\end{figure}

Because the ELMy DIII-D pedestal is not TCP limited, we now calculate the combined ELM and KBM constraints. The ELM constraint is calculated using the ELITE code \cite{Wilson2004}. For each pedestal width and height, we calculate the growth rates corresponding to ideal PBM stability for toroidal mode numbers $n = 5,10,15,20,25$. The growth rates $\gamma$ are normalized to the diamagnetic frequency $\omega_{*i}$. When $\gamma > \omega_{*i}/4 $, the ideal PBM is unstable. In \Cref{fig:width_height_ELM_D3D_162940}(a), we plot $\gamma / \left( \omega_{*i}/4 \right)$ for the fastest growing toroidal mode number. We indicate the contour $\gamma = \omega_{*i}/4$ as the ELM constraint \cite{Snyder2009}. As indicated in \Cref{fig:width_height_ELM_D3D_162940}(a), the equilibrium point is just below the predicted ELM constraint.

The KBM constraint is calculated using the CGYRO and GS2 gyrokinetic codes \cite{Candy2016,Barnes2021}. For each pedestal width and height, gyrokinetic simulations are performed across the pedestal half-width. If the KBM is unstable at all pedestal half-width locations, the pedestal is Gyrokinetic Critical Pedestal unstable \cite{Parisi_2024}. This the kinetic analogue of the nearly-local Ballooning Critical Pedestal criterion used in the EPED model \cite{Snyder2009}.

The combined TCP, KBM, and ELM constraints for DIII-D 162940 are shown in \Cref{fig:width_height_ELM_D3D_162940}(b). The equilibrium is very close to the KBM and ELM constraints, but below the TCP constraint -- as expected for an ELMy pedestal.

\subsection{NSTX}

\begin{figure}[!tb]
    \centering
    \begin{subfigure}[t]{0.48\textwidth}
    \centering
    \includegraphics[width=1.0\textwidth]{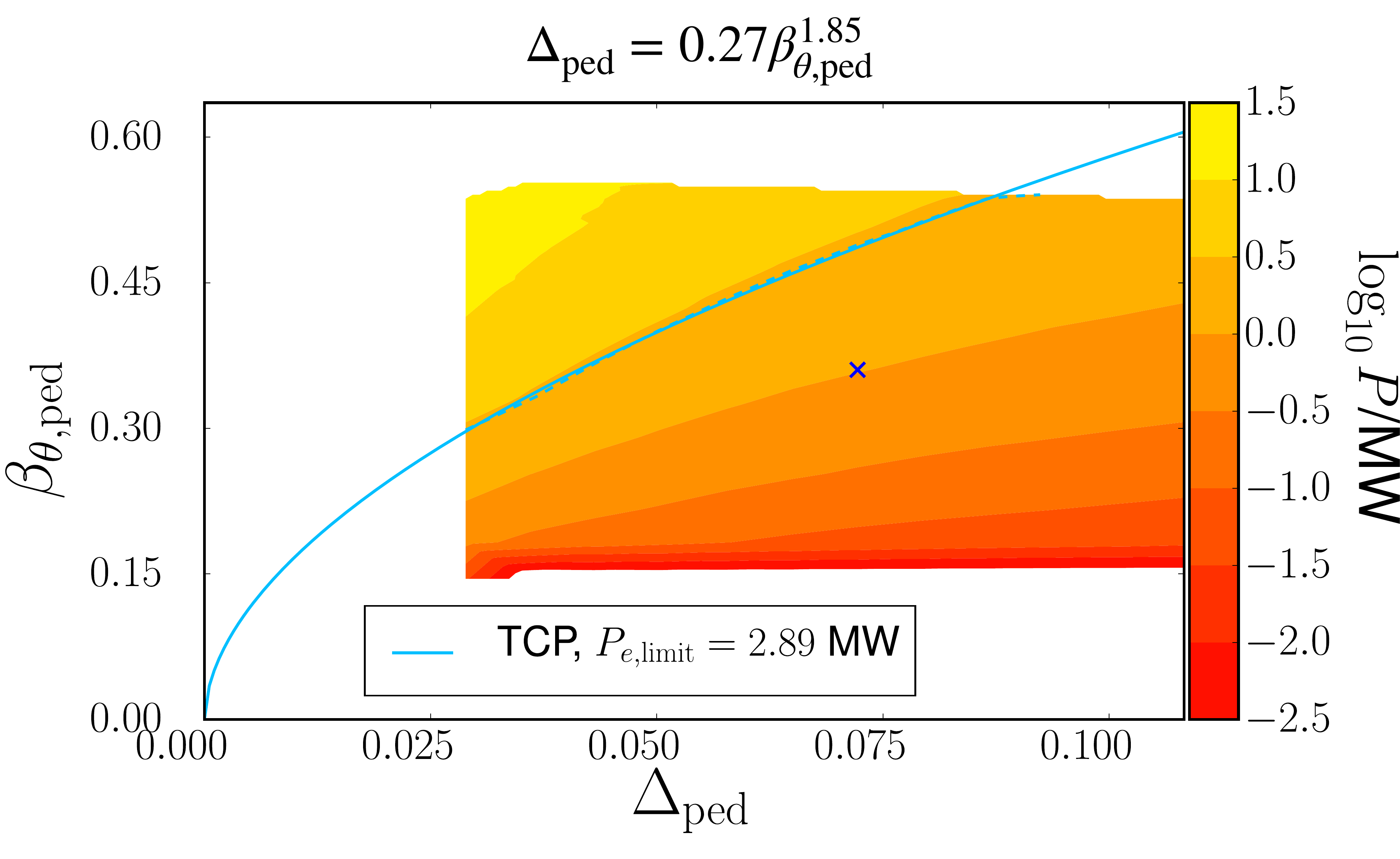}
    \caption{NSTX 132543}
    \end{subfigure}
    ~
    \begin{subfigure}[t]{0.48\textwidth}
    \centering
    \includegraphics[width=1.0\textwidth]{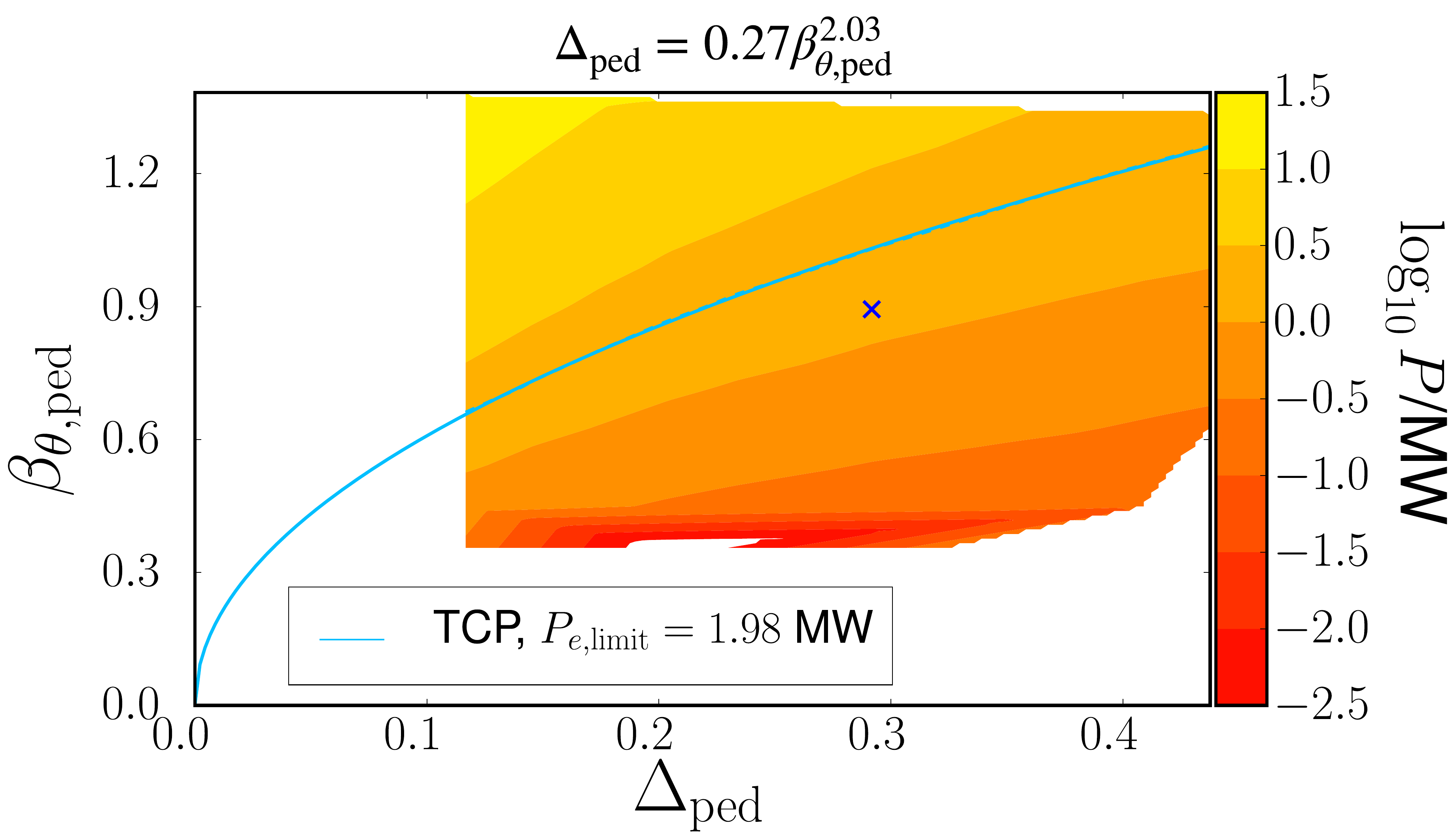}
    \caption{NSTX 132588}
    \end{subfigure}
    ~
    \begin{subfigure}[t]{0.48\textwidth}
    \centering
    \includegraphics[width=1.0\textwidth]{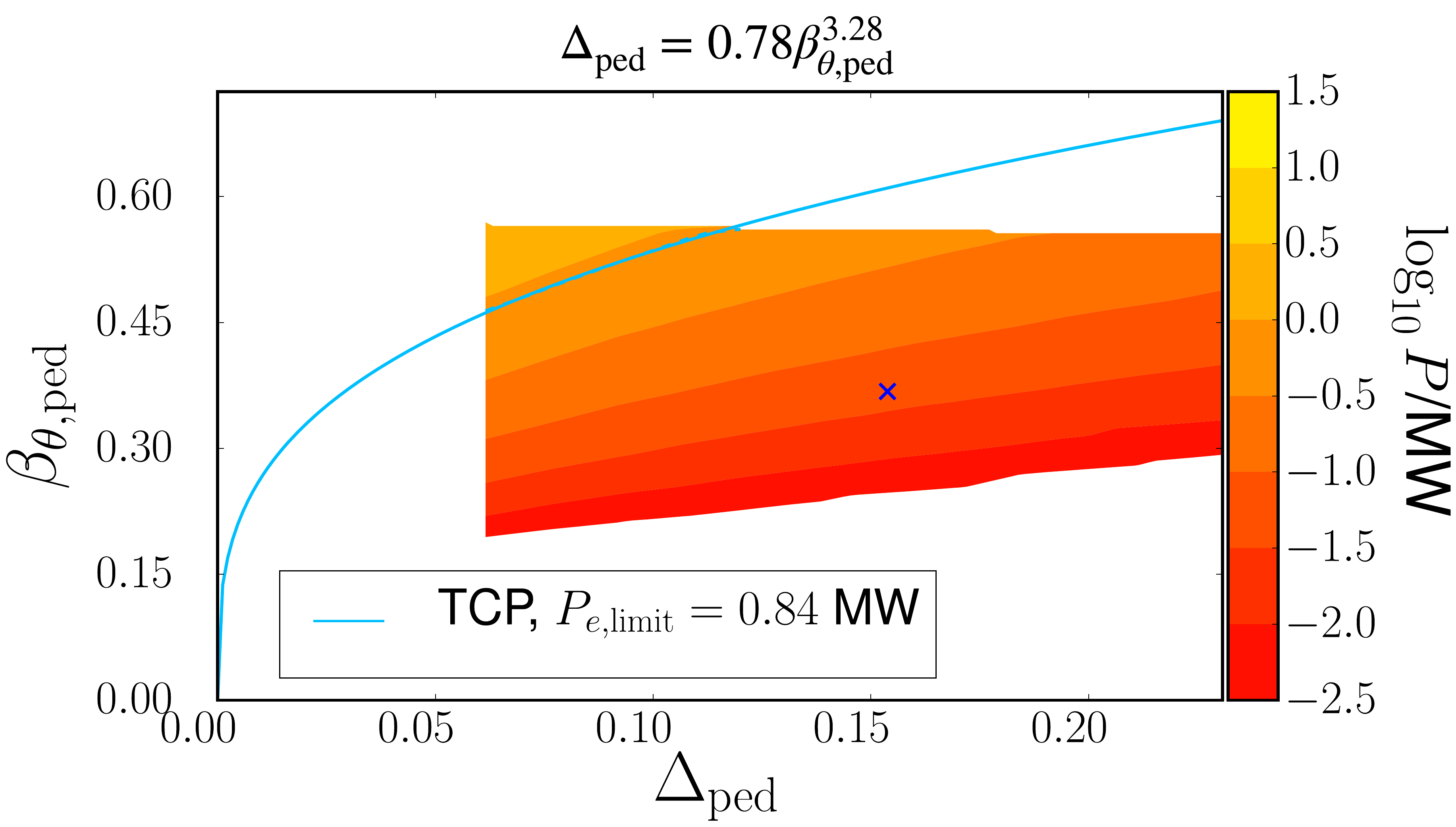}
    \caption{NSTX 129038}
    \end{subfigure}
    \caption{TCP scalings and ETG power $P$ for NSTX discharge 132543, 132588, 129038. The cross marker is the experimental point. All pedestals have constant $n_{e,\mathrm{ped}}$.}
    \label{fig:width_height_NSTX}
\end{figure}

We find the transport width-height scalings for one ELMy (132543, 614ms) and two ELM-free (129038, 600 ms and 132588, 600 ms) NSTX discharges.

The TCP for ELMy discharge 132543 is shown in \Cref{fig:width_height_NSTX}(a). We used a predicted limiting source of $P_{e, \mathrm{limit}} = 2.3$ MW. This gives a TCP of $\Delta_{\mathrm{ped} } = 0.32 \left( \beta_{\theta, \mathrm{ped} } \right)^{1.87} $. See \Cref{app:powerbalance} for a description of how $P_{e, \mathrm{limit}}$ is calculated for this discharge from \Cref{eq:powerbalance} and experimental parameters. According to \Cref{fig:width_height_NSTX}(a), the experimental point has $P = 1.1$ MW. Steepening the pedestal gradients an additional 25\% would cause this pedestal to be ETG transport-limited, but this discharge was already ELMing.

In \Cref{fig:width_height_NSTX}(b), we show the TCP for ELM-free NSTX discharge 132588, $\Delta_{\mathrm{ped} } = 0.27 \left( \beta_{\theta, \mathrm{ped} } \right)^{2.03} $. Given that this is an ultra-wide ELM-free pedestal not originally included in the fitting database for the heat flux formula \cite{Hatch2024}, it is unclear how well $q_e$ in \Cref{eq:heatflux2} applies. Notably, though, the ETG power $P = 1.5$ MW is close to $P_{e, \mathrm{limit}} = 2.0$ MW for this case. Therefore, it is plausible that ETG limits the temperature pedestal at this time for 132588. In \cite{Parisi_2024c} an alternative explanation was provided: KBM and $\mathbf{ E} \times \mathbf{ B}$ constraints intersected roughly at the equilibrium point. Here, a transport-limited pedestal could provide another explanation, although more detailed investigated is required.

In \Cref{fig:width_height_NSTX}(c), we plot the TCP for a second ELM-free NSTX discharge, 129038, $\Delta_{\mathrm{ped} } = 0.78 \left( \beta_{\theta, \mathrm{ped} } \right)^{3.28} $. Unlike 132588, for 129038 the ETG power is over an order of magnitude below $P_{\mathrm{e} } = 0.84$ MW, indicating that the ETG power is highly unlikely to limit this pedestal. We speculate that transport from other gyrokinetic instabilities or a KBM threshold \cite{Parisi_2024b} limits this pedestal.

\subsection{MAST-U}

We find the transport width-height scalings for two MAST-U discharges. MAST-U 48339 is an ELMy discharge at 600ms \cite{Imada2024} and MAST-U 49463 is an ELMy discharge at 600ms that scanned in triangularity throughout the discharge. 49463 subsequently became ELM-free at moderately negative triangularity \cite{Nelson2024b}. At 600ms, 49463 has weakly negative triangularity and is still ELMy.

\begin{figure}[!tb]
    \centering
    \begin{subfigure}[t]{0.48\textwidth}
    \centering
    \includegraphics[width=1.0\textwidth]{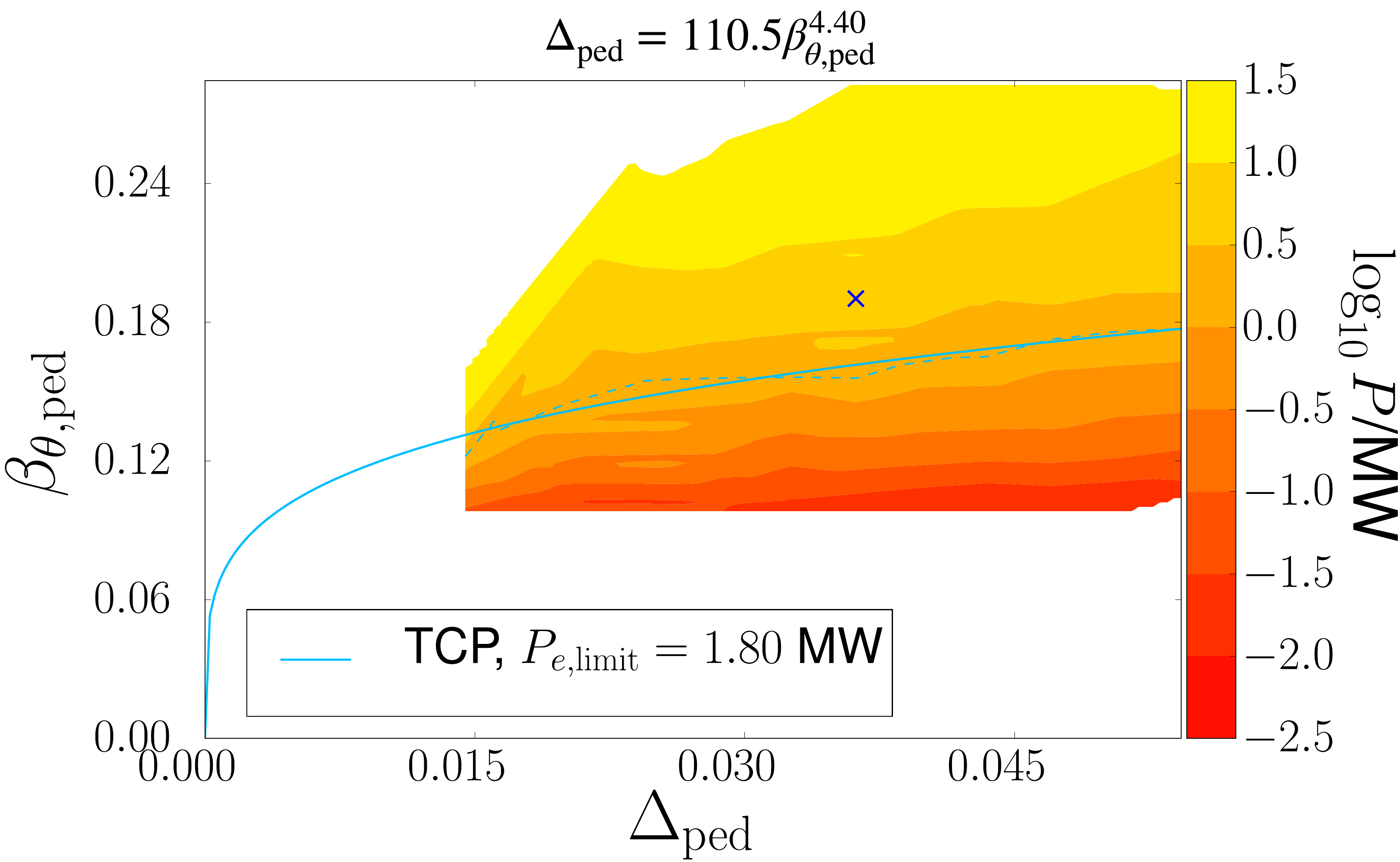}
    \caption{MAST-U 48339}
    \end{subfigure}
    ~
    \begin{subfigure}[t]{0.48\textwidth}
    \centering
    \includegraphics[width=1.0\textwidth]{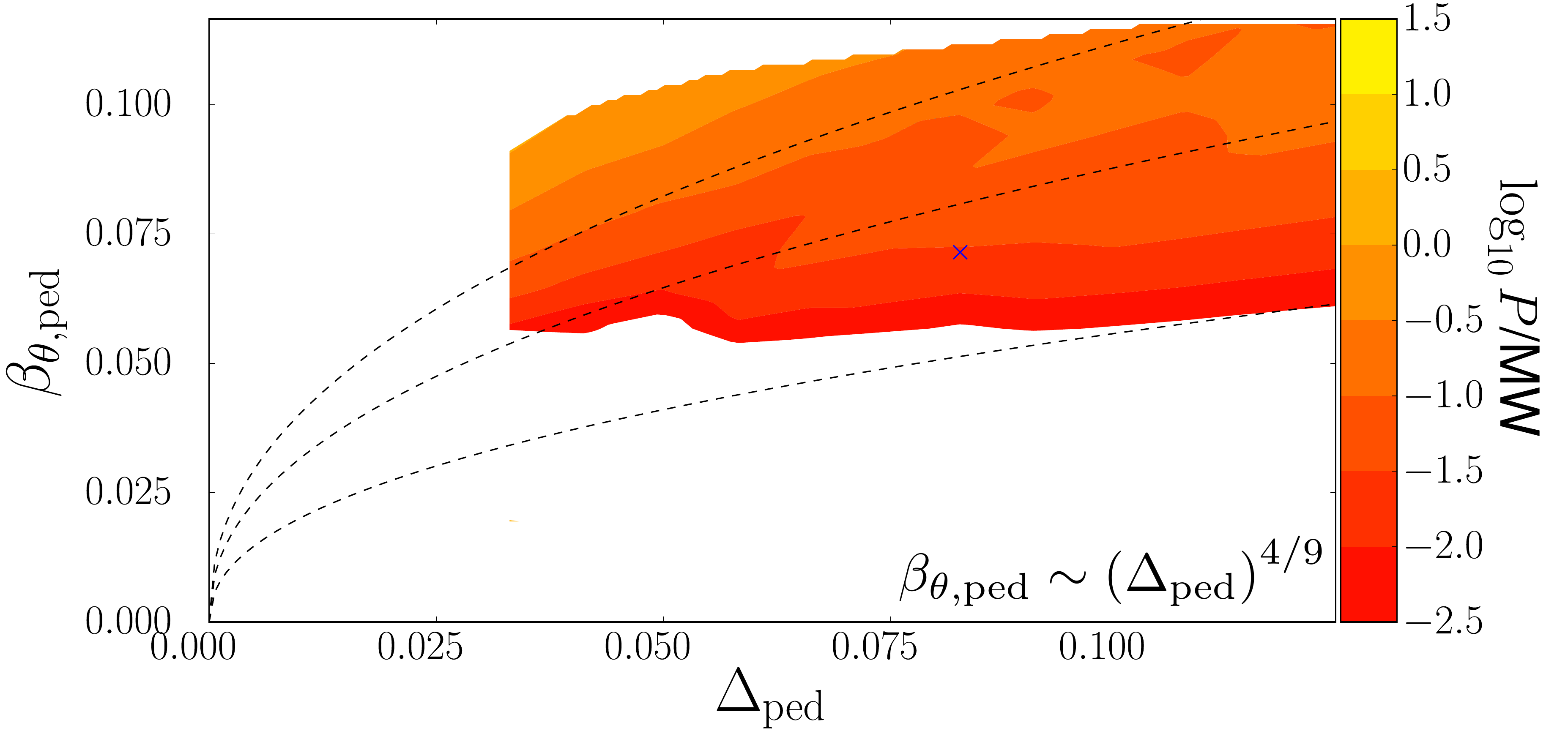}
    \caption{MAST-U 49463.}
    \end{subfigure}
    \caption{TCP scalings and ETG power $P$ for MAST-U 48339 and 49463. The cross marker is the experimental point. All pedestals have constant $n_{e,\mathrm{ped}}$}
    \label{fig:width_height_MASTU}
\end{figure}

\begin{figure}[!tb]
    \begin{subfigure}[t]{0.38\textwidth}
    \centering
    \includegraphics[width=1.0\textwidth]{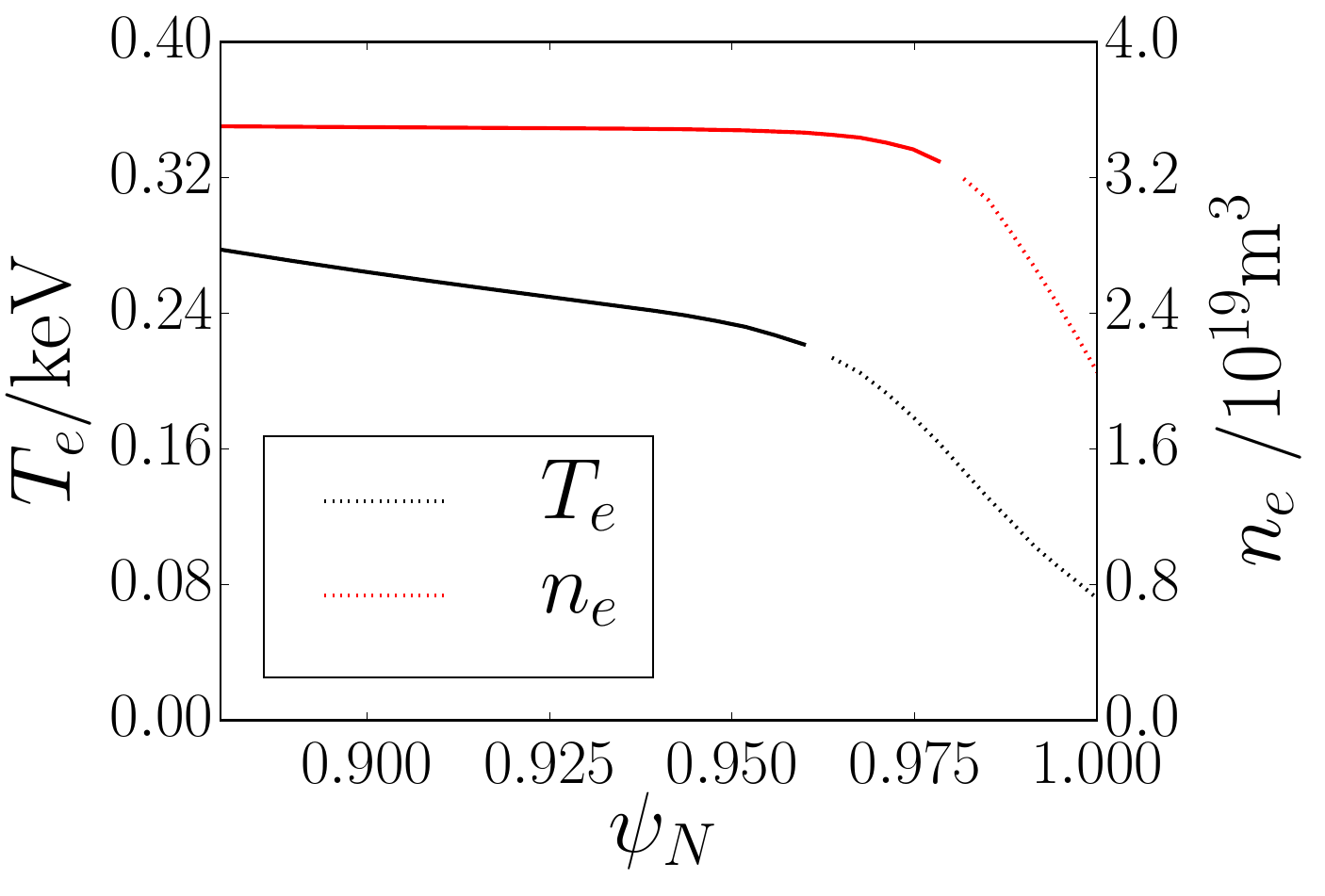}
    \caption{MAST-U 48339}
    \end{subfigure}
    ~
    \begin{subfigure}[t]{0.38\textwidth}
    \centering
    \includegraphics[width=1.0\textwidth]{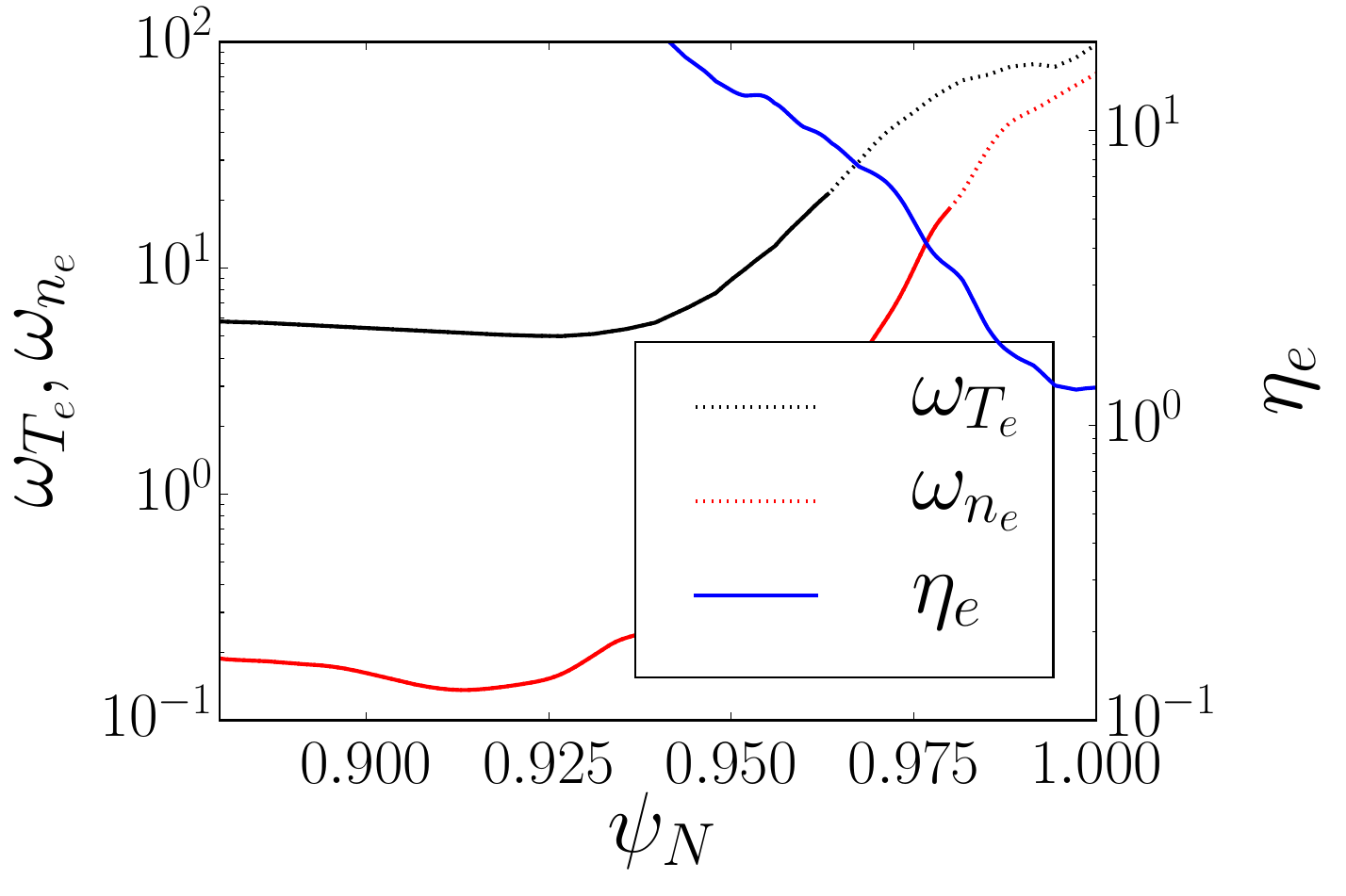}
    \caption{MAST-U 48339}
    \end{subfigure}
    ~
    \begin{subfigure}[t]{0.38\textwidth}
    \centering
    \includegraphics[width=1.0\textwidth]{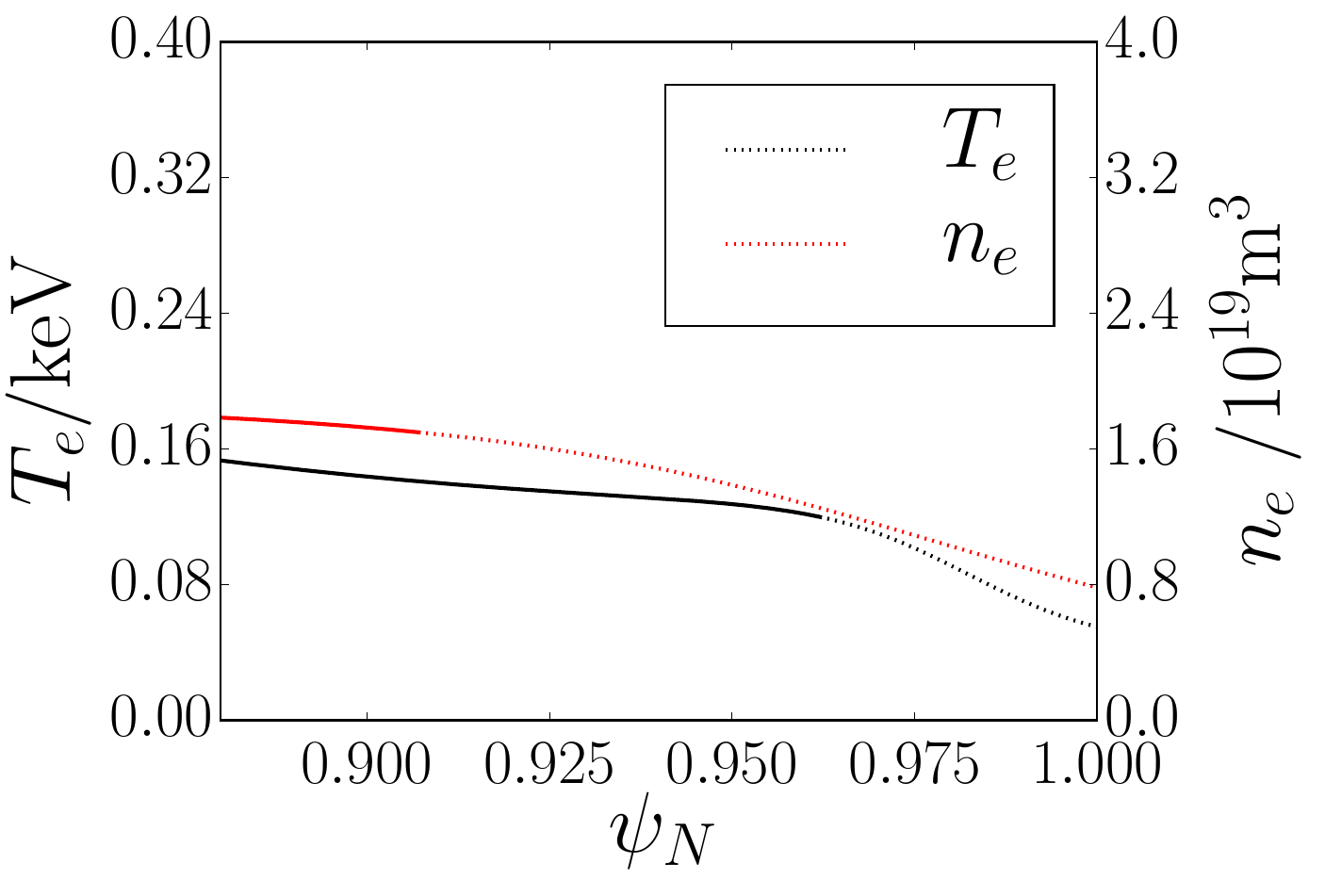}
    \caption{MAST-U 49463}
    \end{subfigure}
    ~
    \begin{subfigure}[t]{0.38\textwidth}
    \centering
    \includegraphics[width=1.0\textwidth]{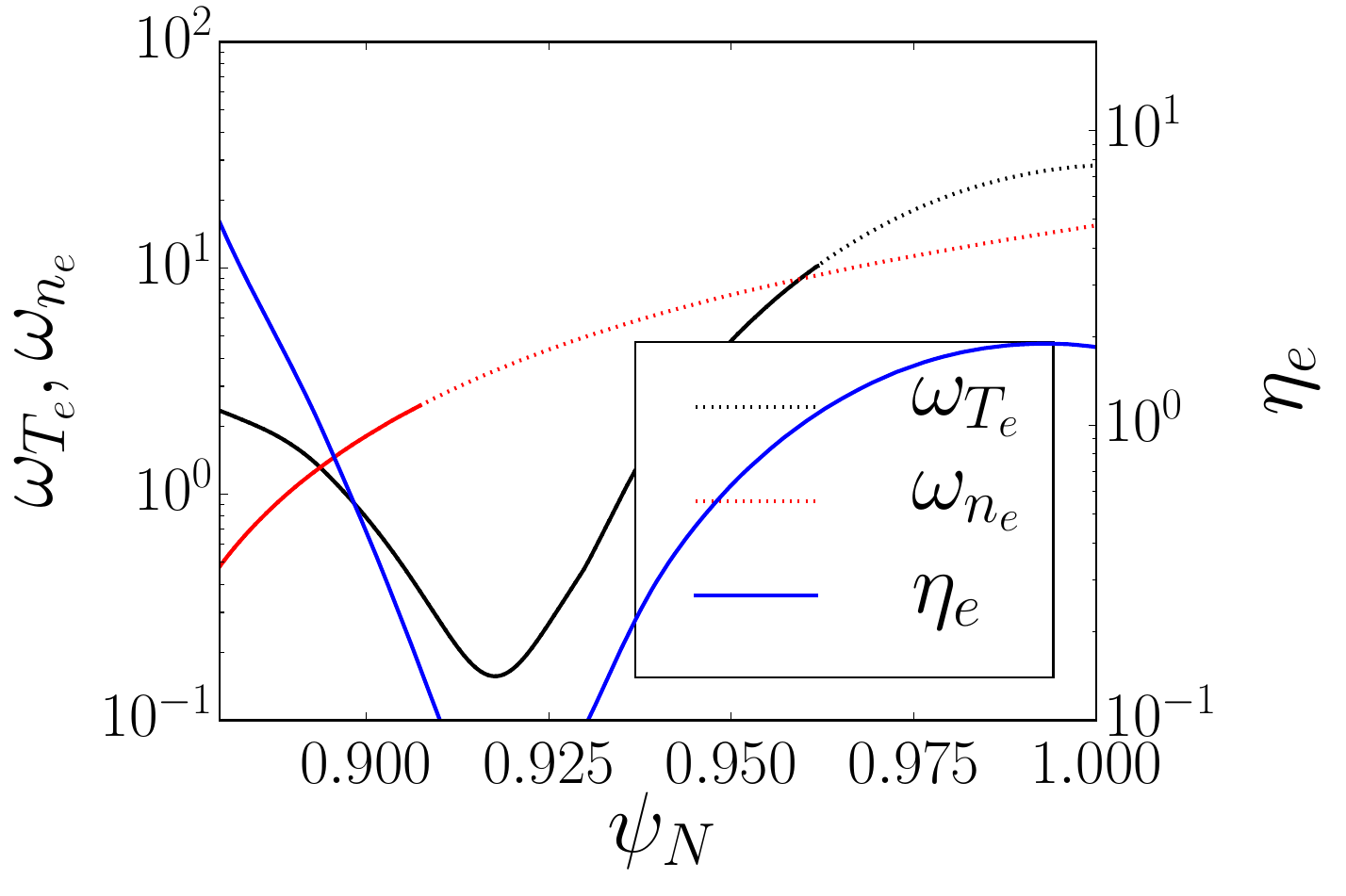}
    \caption{MAST-U 49463}
    \end{subfigure}
    \caption{Density and temperature profiles, and their associated gradients for MAST-U 48339 (a), (b), and 49463 (c), (d). The density pedestals for the two discharges have different relative positions to the temperature pedestals. Dotted lines indicate points within density and temperature pedestal.}
    \label{fig:width_height_MASTU2}
\end{figure}

The TCP for 48339 is shown in \Cref{fig:width_height_MASTU}(a), $\Delta_{\mathrm{ped} } = 110.50 \left( \beta_{\theta, \mathrm{ped} } \right)^{1.40} $. Surprisingly, the predicted ETG power of the experimental equilibrium -- indicated with the cross marker -- is high, $5 = 6.0$ MW. Given that TRANSP \cite{Pankin2024} runs indicate that $P_{e, \mathrm{limit}} = 1.80$ MW, it is extremely likely that $P = 5.0$ MW is an overestimate. To see this, consider the temperature and density fits to profiles for the experimental equilibrium, shown in \Cref{fig:width_height_MASTU2}(a) -- the temperature pedestal is wider than the density pedestal, and is also shifted radially inwards. The relative radial inward shift of $T_e$ has a significant impact on the density and temperature gradients, shown in \Cref{fig:width_height_MASTU2}(b). Towards the pedestal top at $\psi_N \simeq 0.97$, the temperature gradients become very large -- $\omega_{T_e} \simeq 60$ and $\eta_e \simeq 4$. This creates very large fluxes. As discussed in \cite{Hatch2024}, the slab ETG heat flux model may be unsuitable for the pedestal top. This is because toroidal ETG is expected to be a significant additional mechanism, and because of the effects of significantly shifted density and temperature pedestals, particularly when the temperature pedestal is radially inward of the density pedestal. There is an additional effect of $q_{\mathrm{gB} }$ being larger near the pedestal top.

In \Cref{fig:width_height_MASTU}(b), we show the predicted power for MAST-U 49463, which is extremely low. At the experimental point, the ETG power is negligible, $P \simeq 0.03$ MW. One explanation for this is that unlike discharge 48339, discharge 49463 has an opposite shift of the temperature and density pedestals -- the density pedestal is radially inward of the temperature pedestal. This has a significant impact on the gradients: \Cref{fig:width_height_MASTU2}(d) shows how $\omega_{T_e} < 7$ and $\eta_e < 2$ for all locations across the pedestal. Therefore, given that the temperature pedestal is relatively narrow and shifted outward radially relative to the density pedestal, it is plausible that the ETG power in 49463 is orders of magnitude smaller than 48339.

\section{ELM-Free Operation} \label{sec:ELMfree}

\begin{figure}[!tb]
    \centering
    \begin{subfigure}[t]{0.4\textwidth}
    \centering
    \includegraphics[width=1.0\textwidth]{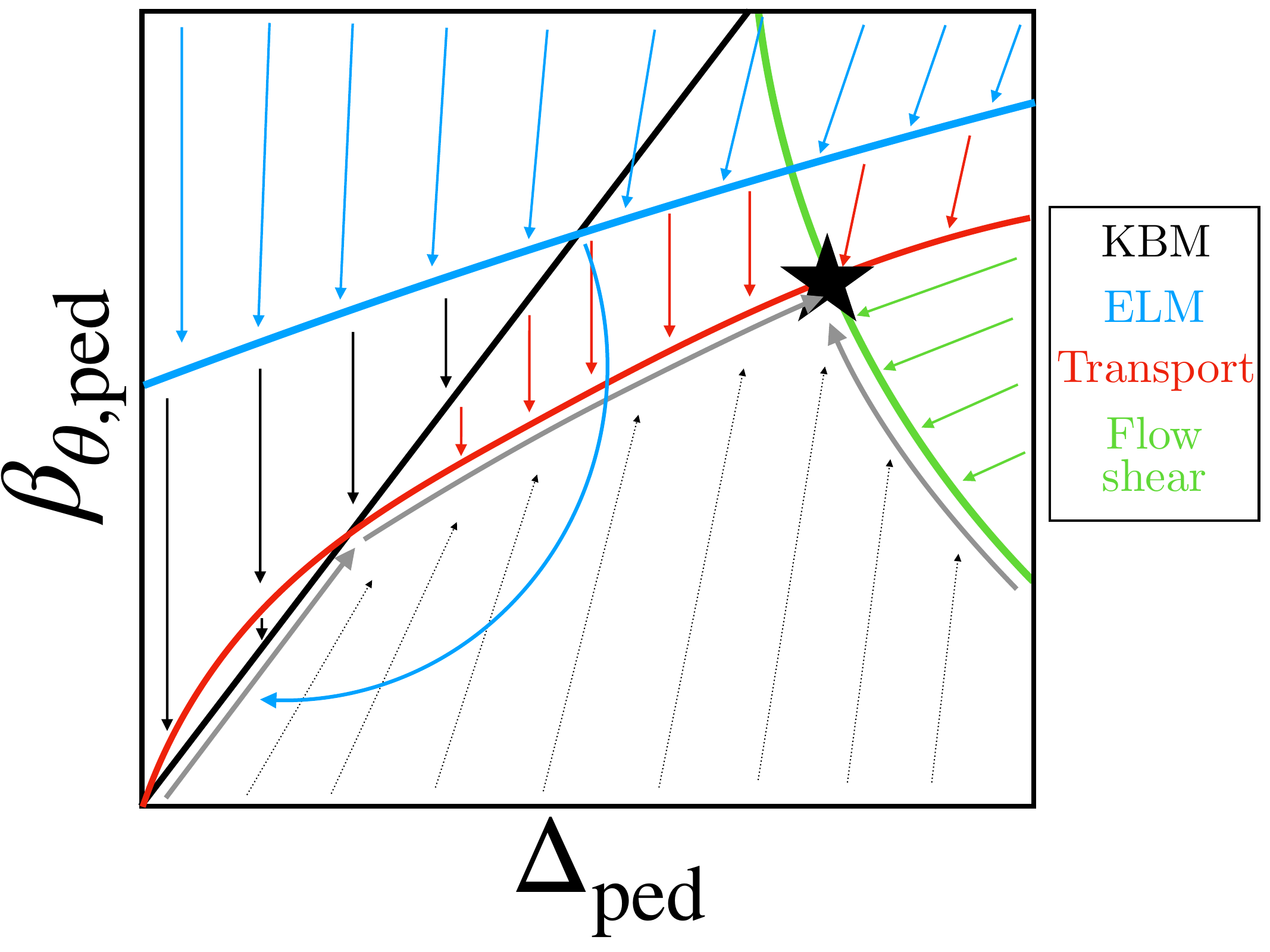}
    \caption{}
    \end{subfigure}
    ~
    \begin{subfigure}[t]{0.4\textwidth}
    \centering
    \includegraphics[width=1.0\textwidth]{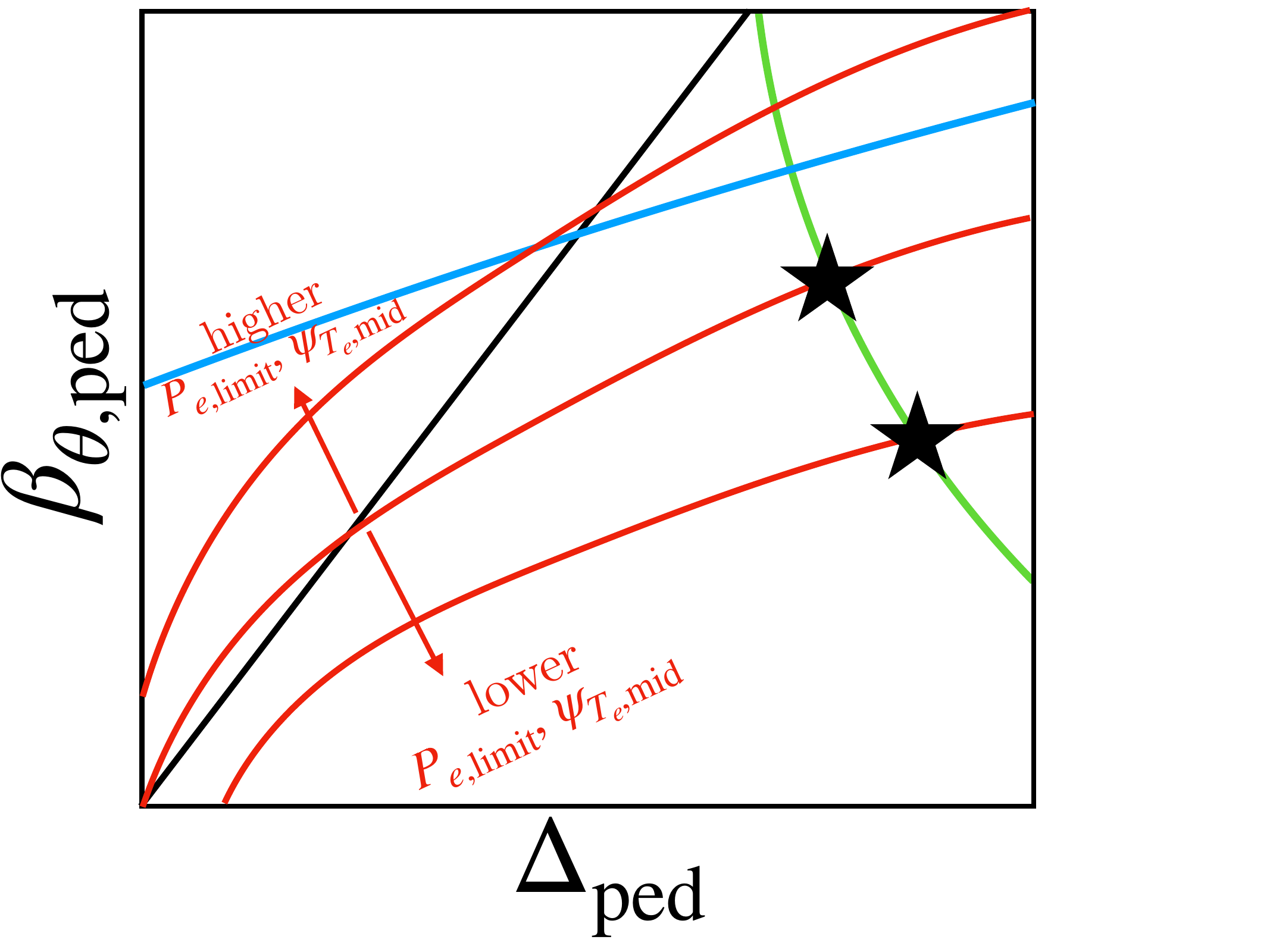}
    \caption{}
    \end{subfigure}
    ~
    \begin{subfigure}[t]{0.4\textwidth}
    \centering
    \includegraphics[width=1.0\textwidth]{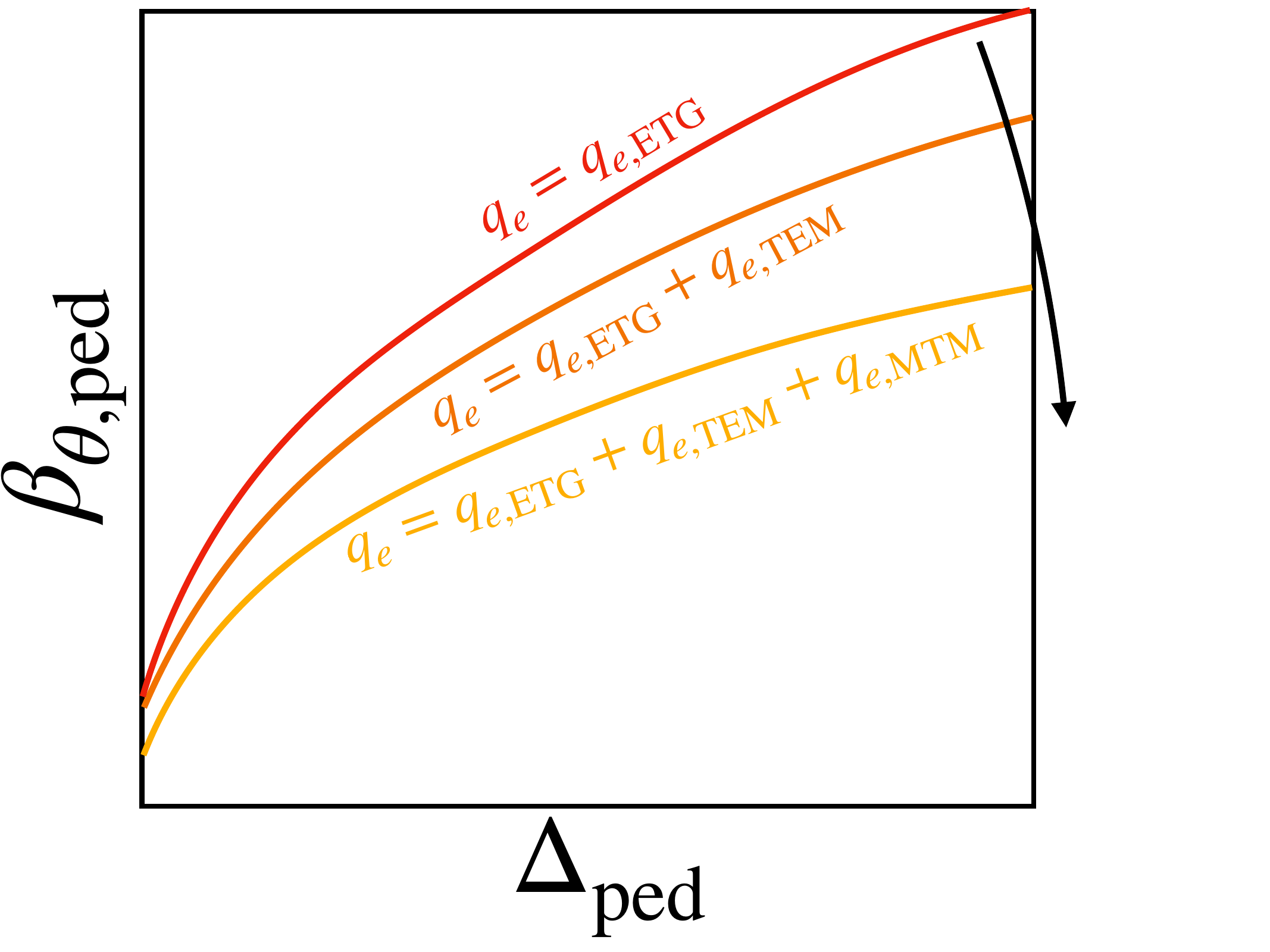}
    \caption{}
    \end{subfigure}
    \caption{Pedestal phase diagrams. Stars indicate stable points. (a) vector diagram showing phase space trajectories. (b) the effect of shifting transport constraint by changing $P_{e,\mathrm{limit} }$ or $\psi_{T_e,\mathrm{mid} }$. (c) Schematic of adding more electron heat transport mechanisms to TCP width-height scaling.}
    \label{fig:width_height_flow_charts}
\end{figure}

In this section, we discuss prospects for ELM-free operation in transport-limited pedestals.

A noteworthy outcome of this paper is that the ELM scaling of width tends to have a weaker scaling with $\beta_{\theta,\mathrm{ped} }$ than for ETG transport if most of the pedestal buildup is via the temperature: the PBM width-height scaling typically satisfies $\Delta_{\mathrm{ped} } \sim \left( \beta_{\theta, \mathrm{ped} }\right)^{4/3}$ \cite{Snyder2009}, whereas results for NSTX, MAST-U, and DIII-D at fixed $n_{e,\mathrm{ped} }$ here had $\Delta_{\mathrm{ped} } \sim \left( \beta_{\theta, \mathrm{ped} }\right)^{\gamma}$ with $\gamma \gtrsim 2$. Additionally, the ELM constraint often has a positive pressure offset at $\Delta_{\mathrm{ped} } = 0$ \cite{Snyder2009b}. The weaker ELM scaling of width with $\beta_{\theta,\mathrm{ped} }$ is promising because it means that the ELM and TCP scalings may never intersect, meaning wider pedestals on a transport-limited trajectory may never hit an ELM limit.

We therefore arrive at the pedestal space diagram in \Cref{fig:width_height_flow_charts}(a). The arrows of the same color as the KBM, ELM, transport, or flow shear constraints show the direction that these constraints tend to move the pedestal. The ELM constraint is more complicated because it is a macroscopic instability that significantly reduces the pedestal width and height.

There still remains the issue of what saturates pedestal growth. For a pedestal equilibrium point, there must be two constraints that have have $\beta_{\theta, \mathrm{ped}}$ scalings with $\Delta_{\mathrm{ped} }^{c_0}$, where the exponents $c_0$ for the two scalings have opposite signs, otherwise the pedestal will continue to grow. One such mechanism is the $\mathbf{ E} \times \mathbf{ B}$ flow shear. There is only one stable equilibrium point on this diagram -- marked with a star. This is the intersection of the transport and flow shear constraints.

Another property of steady-state transport-limited pedestals is that changes in the sources leads to different trajectories. Hence, the sources are an actuator for the pedestal width-height relation, and might be used to change the proximity to the ELM -- this is shown in \Cref{fig:width_height_flow_charts}(b). The larger the electron sources, the closer the equilibrium point -- marked by stars in \Cref{fig:width_height_flow_charts}(b) -- is to the ELM limit. Another actuator for the pedestal height through the TCP is the relative shift of the temperature to the density pedestal, as indicated in \Cref{fig:width_height_flow_charts}(b).

\section{Heat and Particle Transport Model} \label{sec:combined_particle_heat_model}

In this section, we expand the pedestal width-height scaling model based loosely on the ETG and KBM model in \cite{Guttenfelder2021}. First, as in \cite{Guttenfelder2021} we propose that the particle and heat diffusivity ratios for ETG and KBM are constrained to be a constant value,
\begin{equation}
\frac{D_e}{\chi_e} \bigg{|}_\mathrm{ETG} = 0.02, \;\;\; \frac{D_e}{\chi_e} \bigg{|}_\mathrm{KBM} = 1.00.
\end{equation}
Writing the heat and particle fluxes in diffusive form
\begin{equation}
\mathbf{\Gamma}_e = - D_e \nabla n_e, \;\;\; \mathbf{q}_e = - \chi_e n_e \nabla T_e,
\end{equation}
gives an expression for the particle flux,
\begin{equation}
\Gamma_e = \frac{D_e}{\chi_e} \frac{1}{T_e \eta_e} q_e. 
\end{equation}
We use the same KBM electron heat flux model as in \cite{Guttenfelder2021},
\begin{equation}
\chi_{e, \mathrm{KBM} } = C_\mathrm{KBM} \left( \alpha - \alpha_\mathrm{crit} \right) \left( \frac{\rho_s^2 c_s}{a} \right),
\end{equation}
where $C_\mathrm{KBM}$ is a constant - here we chose $C_\mathrm{KBM} = 0.6$, $\alpha$ is the Shafranov shift, and $\alpha_\mathrm{crit}$ is the critical $\alpha$ value for KBM instability. While \cite{Guttenfelder2021} assumed that $\alpha_\mathrm{crit}$ is given by the experimental profile, in this work we calculate $\alpha_\mathrm{crit}$ using the ideal infinite-$n$ ballooning boundary for each equilibrium and each flux surface, scaled by $0.7$ \cite{Parisi_2024} In this limited example, we are only interested in the KBM transport up to first stability for an NSTX pedestal -- our model does not capture the effects of second stability  at very steep gradients and low magnetic shear; in this regime, the KBM fluxes could vanish. This would be a straightforward generalization of the current model. A more thorough investigation would also vary $\alpha$ self-consistently to find $\alpha_\mathrm{crit}$ for the KBM (as opposed to the ideal infinite-$n$ ballooning mode) at each flux surface and each equilibrium -- the rescaling of the ideal infinite-$n$ ballooning $\alpha_\mathrm{crit}$ is a rough heuristic based on intuition developed in \cite{Parisi_2024,Parisi_2024b,Parisi_2024c}. Finding $\alpha_\mathrm{crit}$ for the KBM is computationally intensive and far beyond the scope of this paper.

Using the same approach for calculating the limiting particle flux as for the heat flux, a particle-transport-limited pedestal satisfies
\begin{equation}
\frac{\partial n_e}{\partial  t} = \sum_k L_{e,k} - \nabla \cdot \mathbf{\Gamma}_e = 0,
\end{equation}
where $L_{e,k}$ are electron particle sources. In this transport-limited pedestal, the turbulent particle flux $G$ is equal to the net source $G_{\mathrm{limit} }$ at some flux surface in the pedestal,
\begin{equation}
G_e \equiv \Gamma_e S = G_{e,\mathrm{limit} } \equiv \int \sum_k L_{e,k} d V,
\label{eq:particlebalance}
\end{equation}
Assuming that $\sum_k L_{e,k}$ does not vary significantly over the pedestal, when finding $G_{e,\mathrm{limit} }$ we will evaluate the right-hand side of \Cref{eq:particlebalance} close to the separatrix at $\psi_N = 0.98$. Note that the assumption of $\sum_k L_{e,k}$ not varying substantially over the pedestal is less well-justified for particle than for energy, as we show in \Cref{app:particlebalance} -- the quantity $\sum_k L_{e,k}$ can vary by a factor of two over the pedestal due to the strong variation in the wall neutral source. Therefore, by limiting our evaluation of $G_{e,\mathrm{limit} }$ to the $\psi_N = 0.98$ flux surface, we simplify the particle transport model by no longer requiring a model for how the particle sources vary across the pedestal. This is also a significant drawback, as it means that in this model the density pedestal is only limited by the particle flux and source near the separatrix -- there may be indeed be pedestals that are particle transport-limited closer to the pedestal top or mid-pedestal that are not captured by this model.

To summarize, the model equations for the heat flux are
\begin{equation}
q_e = q_{e,\mathrm{ETG}} + q_{e,\mathrm{KBM}},
\end{equation}
where
\begin{equation}
\mathbf{q}_{e,\mathrm{KBM}} = - \chi_{e,\mathrm{KBM} } n_e \nabla T_e.
\end{equation}
We use a different model for $q_{e,\mathrm{ETG}}$ than \cite{Guttenfelder2021} -- we use \Cref{eq:qHatch23} (from \cite{Hatch2024}). The model equations for the particle flux are
\begin{equation}
\Gamma_e = \Gamma_{e,\mathrm{ETG}} + \Gamma_{e,\mathrm{KBM}},
\end{equation}
where the ETG particle flux is
\begin{equation}
\Gamma_{e,\mathrm{ETG}} = \frac{D_e}{\chi_e} \bigg{|}_\mathrm{ETG} \frac{1}{T_e \eta_e} q_{e,\mathrm{ETG} },
\end{equation}
and the KBM particle flux is
\begin{equation}
\Gamma_{e,\mathrm{KBM}} = \frac{D_e}{\chi_e} \bigg{|}_\mathrm{KBM} \frac{1}{T_e \eta_e} q_{e,\mathrm{KBM} }.
\end{equation}
For both the heat and particle fluxes, we assume that the fluxes are source-limited according to \Cref{eq:powerbalance,eq:particlebalance}
\begin{equation}
P_e = P_\mathrm{limit}, \;\;\;\; G_e = G_\mathrm{limit}.
\end{equation}
Similar to previous sections, we find the radial locations where $P_e$ and $G_e$ are a maximum (the radial locations for maximum $P_e$ and $G_e$ may differ) and compare with $P_\mathrm{limit}$ and $G_e = G_\mathrm{limit}$. Because of the lack of a validated algebraic expression for KBM transport, we choose $C_\mathrm{KBM} = 0.6$ to satisfy the constraint $G_e = G_\mathrm{limit}$ for the experimental NSTX equilibrium.

\begin{figure}[!tb]
    \centering
    \begin{subfigure}[t]{0.49\textwidth}
    \centering
    \includegraphics[width=1.0\textwidth]{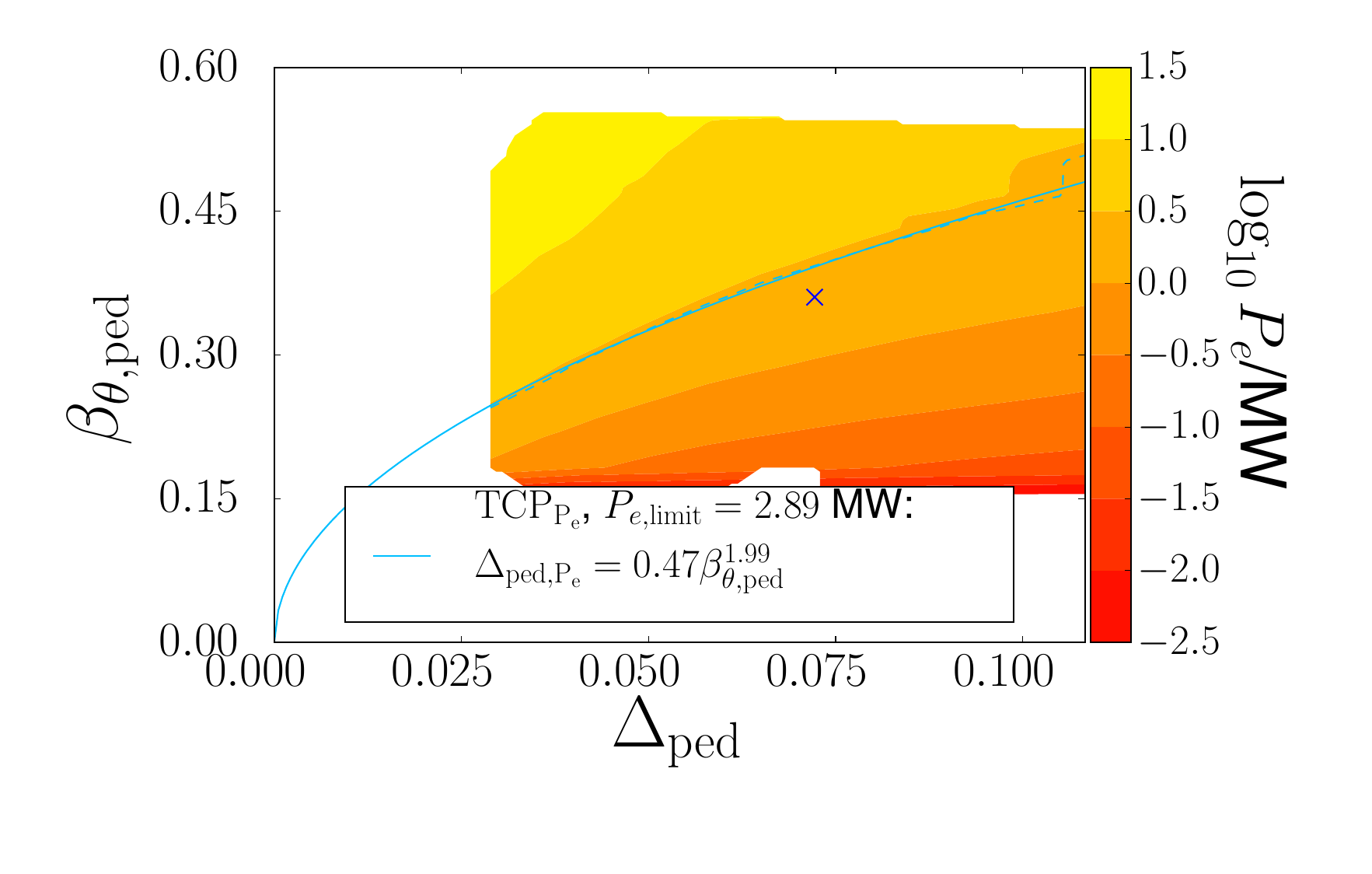}
    \caption{}
    \end{subfigure}
    \begin{subfigure}[t]{0.49\textwidth}
    \centering
    \includegraphics[width=1.0\textwidth]{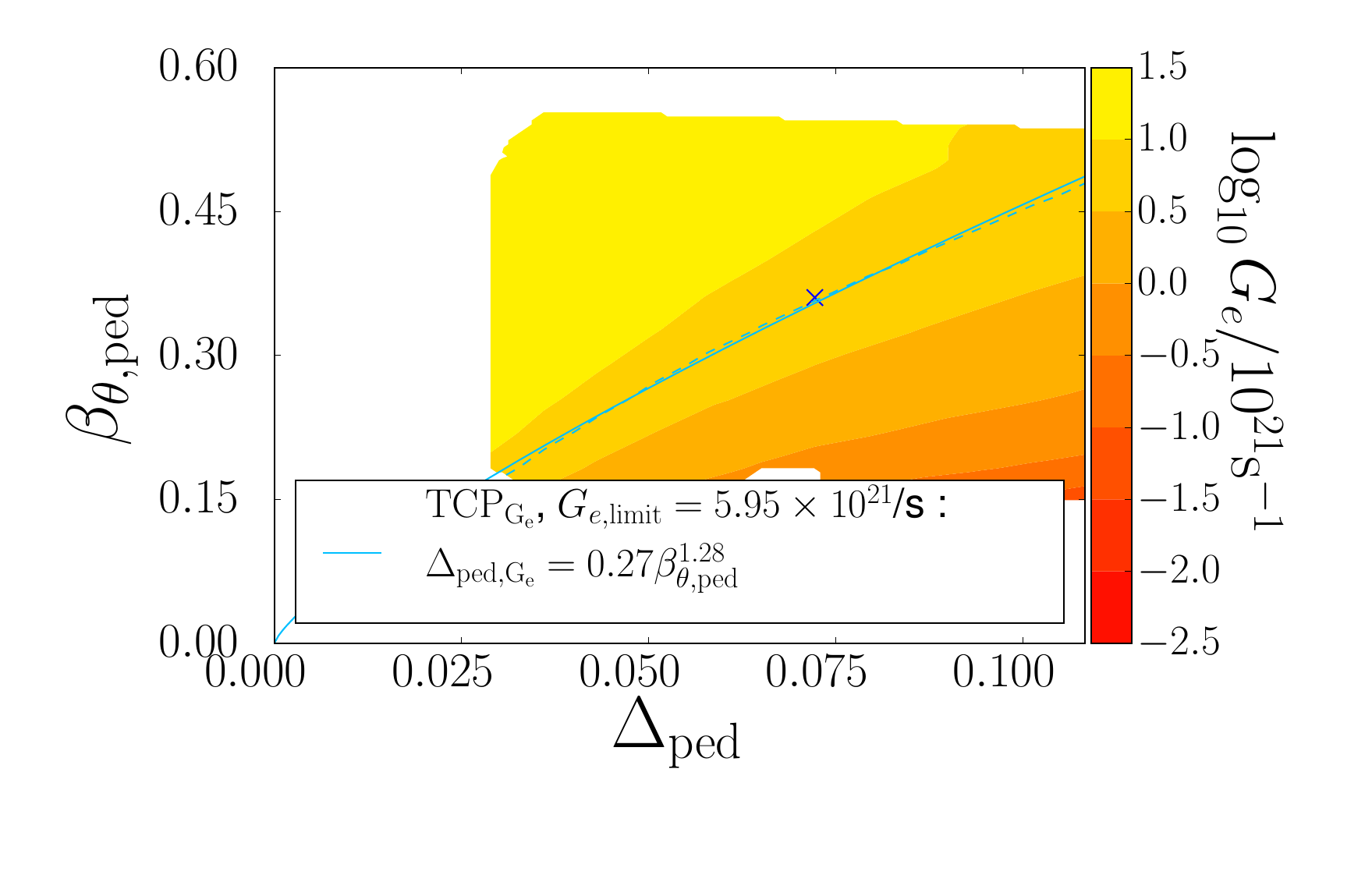}
    \caption{}
    \end{subfigure}
     ~
    \caption{(a) Power and (b) particle transport scalings for NSTX 132543 with a combined ETG and KBM transport model. The pedestal height is varied with fixed $n_{e,\mathrm{ped} }$}
    \label{fig:width_height_scalings_KBM_ETG_model}
\end{figure}

\begin{figure}[!tb]
    \centering
    \begin{subfigure}[t]{0.49\textwidth}
    \centering
    \includegraphics[width=1.0\textwidth]{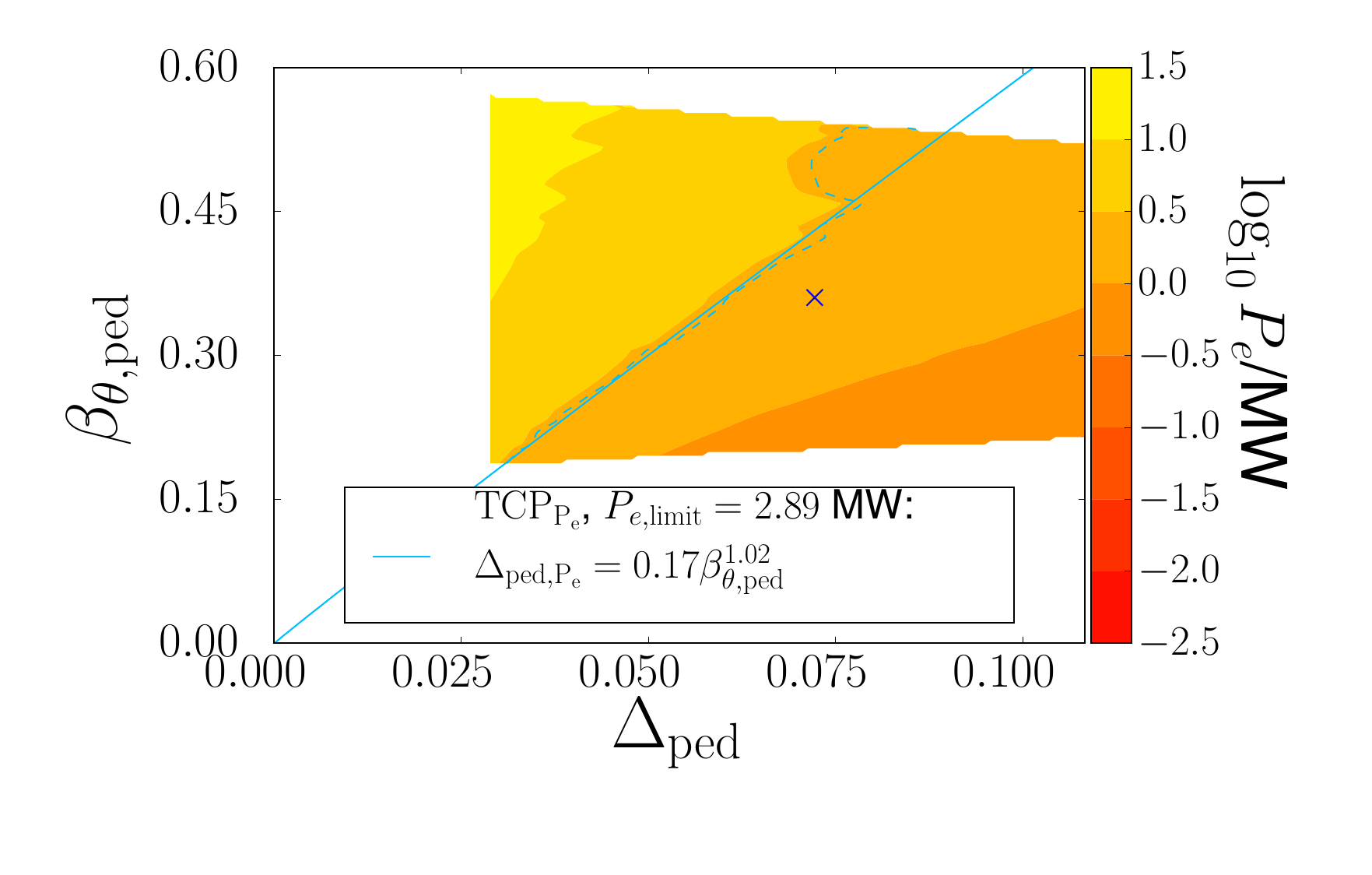}
    \caption{}
    \end{subfigure}
    \begin{subfigure}[t]{0.49\textwidth}
    \centering
    \includegraphics[width=1.0\textwidth]{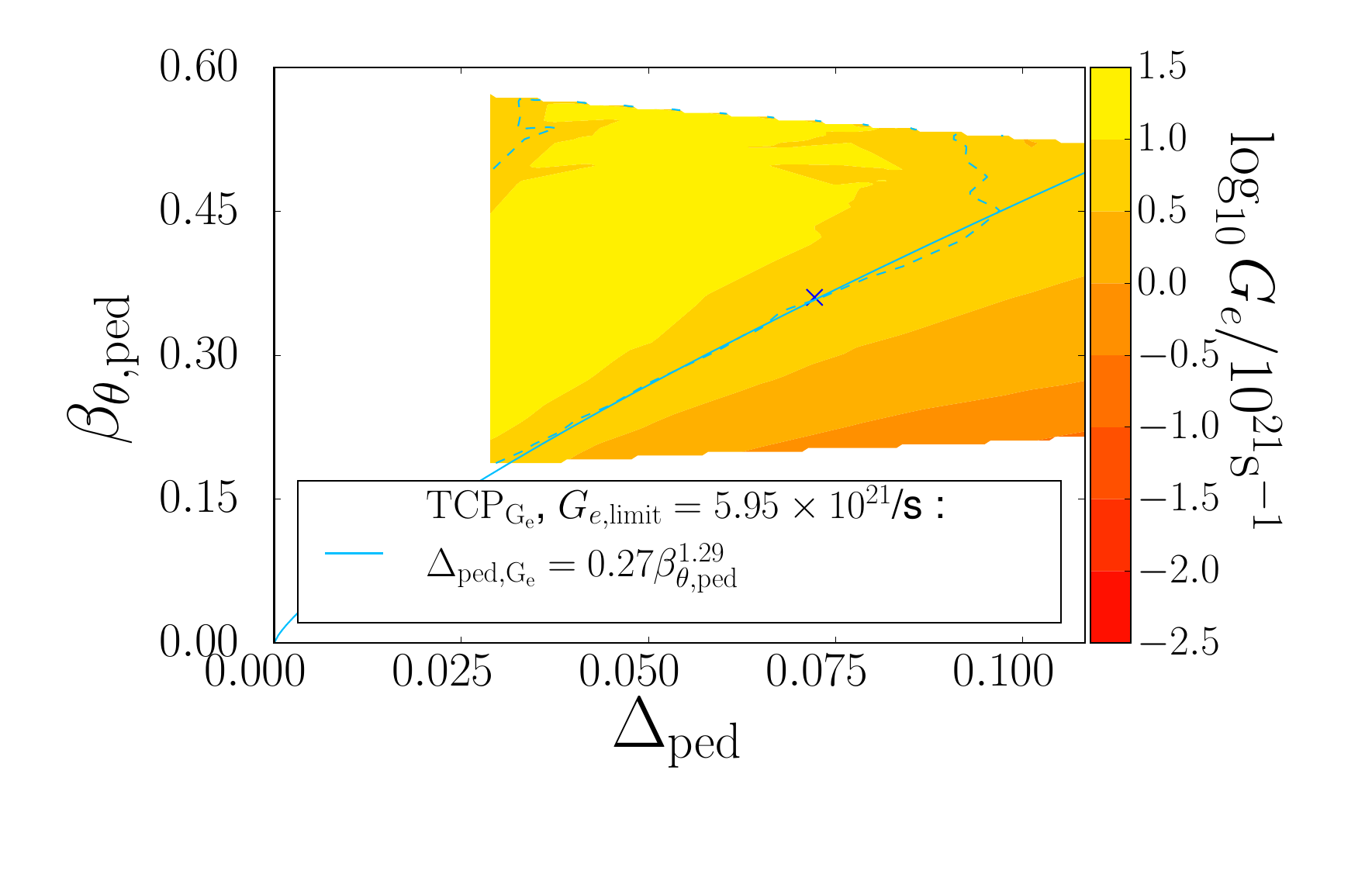}
    \caption{}
    \end{subfigure}
     ~
    \caption{(a) Power and (b) particle transport scalings for NSTX 132543 with a combined ETG and KBM transport model. The pedestal height is varied with fixed $T_{e,\mathrm{ped} }$}
    \label{fig:width_height_scalings_KBM_ETG_model_fixedT}
\end{figure}

The trajectories of $P_e (\Delta_\mathrm{ped}, \beta_{\theta,\mathrm{ped} }) = P_\mathrm{limit}$ and $G_e (\Delta_\mathrm{ped}, \beta_{\theta,\mathrm{ped} }) = P_\mathrm{limit}$ are referred to as $\mathrm{TCP}_{\mathrm{P_e} }$ and $\mathrm{TCP}_{\mathrm{G_e} }$. These trajectories, shown in \Cref{fig:width_height_scalings_KBM_ETG_model} for the NSTX discharge 132543 analyzed earlier in this work, reveal some curious features. First, the power-constrained scaling $\mathrm{TCP}_{\mathrm{P_e} }$ has a quadratic scaling
\begin{equation}
\Delta_{\mathrm{ped,P_e} } = 0.47 \beta_{\theta, \mathrm{ped} }^{1.99},
\end{equation}
whereas the particle-constrained scaling $\mathrm{TCP}_{\mathrm{G_e} }$ has a scaling
\begin{equation}
\Delta_{\mathrm{ped,G_e} } = 0.27 \beta_{\theta, \mathrm{ped} }^{1.28}.
\end{equation}
The scaling for $\Delta_{\mathrm{ped,G_e} }$ is closer to NSTX experimental measurements \cite{Diallo2013} and gyrokinetic modeling \cite{Parisi_2024}. Given that the electron particle source is typically weak \cite{Kotschenreuther2019} and there are few other turbulent modes apart from KBM to cause significant particle transport, it may not be surprising that the particle transport scaling associated with KBM + ETG is in reasonable agreement with experiment. It is also noteworthy that the particle TCP constraint occurs at lower $\beta_{\theta,\mathrm{ped} }$ for narrower pedestals -- since the NSTX 132543 is near the ELM-cycle top for the timeslice we analyze, this indicates that particle transport is more limiting than heat transport for constraining this particular pedestal. Finally, it is important to note that in \Cref{fig:width_height_scalings_KBM_ETG_model} we varied pedestal height at fixed $n_{e,\mathrm{ped} }$. In \Cref{fig:width_height_scalings_KBM_ETG_model_fixedT}, we vary pedestal height at fixed $T_{e,\mathrm{ped} }$. Curiously, the particle-constrained scaling $\mathrm{TCP}_{\mathrm{G_e} }$ is basically unchanged but the power-constrained scaling $\mathrm{TCP}_{\mathrm{P_e} }$ now has a linear rather than quadratic scaling. For both fixed $n_{e,\mathrm{ped} }$ and fixed $T_{e,\mathrm{ped} }$, the particle TCP constraint occurs at lower $\beta_{\theta,\mathrm{ped} }$ for narrower pedestals.

In summary, we used a reduced transport model to calculate the pedestal width-height scaling for a pedestal limited by electron heat or particle transport using arising from ETG and KBM turbulence. We emphasize that this is a low-fidelity demonstration of concept, and that more much work is required to validate this approach.

\section{Model Extensions} \label{sec:modelextensions}

\begin{figure}[!tb]
    \centering
    \begin{subfigure}[t]{0.4\textwidth}
    \centering
    \includegraphics[width=1.0\textwidth]{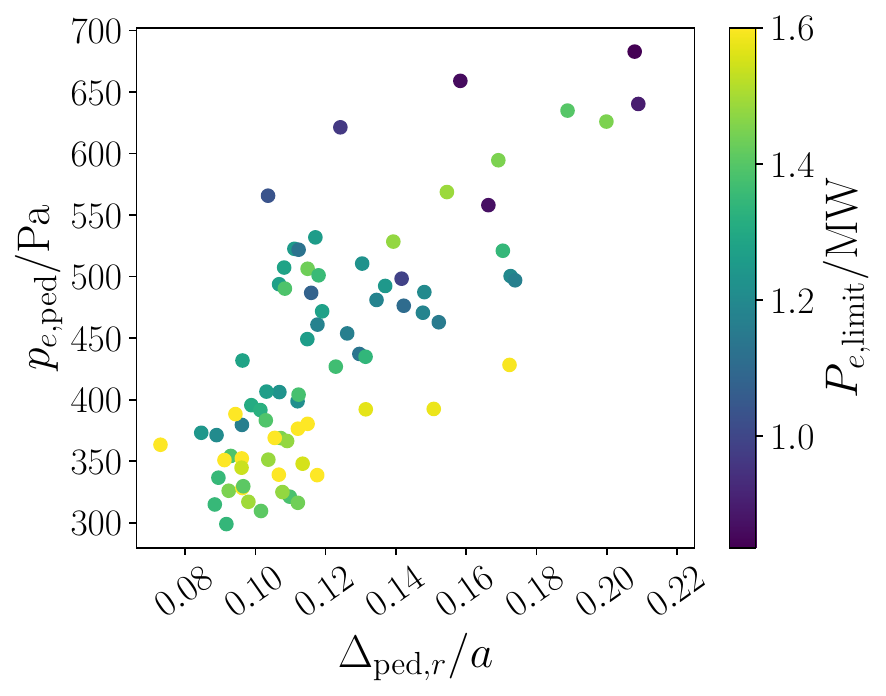}
    \caption{}
    \end{subfigure}
    \begin{subfigure}[t]{0.4\textwidth}
    \centering
    \includegraphics[width=1.0\textwidth]{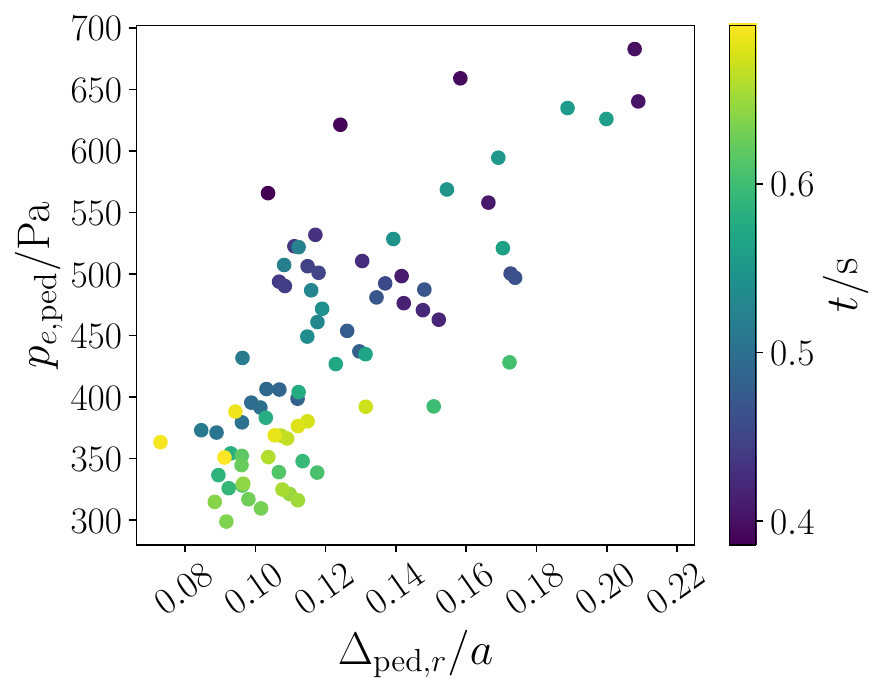}
    \caption{}
    \end{subfigure}
     ~
    \caption{For MAST-U 49463, total electron heat source $P_{e,\mathrm{limit} }$ (a) and (b) discharge time $t$ versus pedestal electron height $p_{e,\mathrm{ped} }$ and $\Delta_{\mathrm{ped} ,r}/a$. Here, $\Delta_{\mathrm{ped} ,r}/a$ is the pedestal width in meters divided by minor radius $a$.}
    \label{fig:Pesource_widthheightspace}
\end{figure}

In this section, we discuss possible model extensions and outstanding questions.

In this paper, we initially considered electron heat transport resulting from ETG-driven turbulence. The reasons for this are (1) conceptual simplicity and (2) recent availability of analytic formulae for ETG heat transport in the pedestal \cite{Hatch2022a,Hatch2024,Farcas2024}. However, a more useful and more falsifiable reduced model would include (a) transport from additional instabilities, (b) in all relevant channels: density, momentum, and heat, (c) for both ion and electron species. There are limited examples of pedestals where ETG transport accurately predicts the electron temperature profile, for example, a JET H-mode \cite{Field2023} and an Alcator C-Mod I-mode \cite{Hatch2024}. Therefore, the reduced model in this paper might be testable on such discharges.

One could extend the electron heat transport model to include additional instabilities as implemented in many transport existing models \cite{Hawryluk1981,Jardin1993,Waltz1998,Bateman1998,Pereverzev2002,Staebler2007,Belli2008b,Rafiq2013}. The main challenge of including more transport mechanisms is the lack of validated reduced algebraic models for turbulent transport mechanisms in the pedestal. A total electron heat flux expression would have the form $q_{e,\mathrm{total} } = q_{e,\mathrm{ETG} } + q_{e,\mathrm{ITG} } + q_{e,\mathrm{TEM} }+ q_{e,\mathrm{MTM} }$, where ion-temperature-gradient (ITG), trapped-electron-mode (TEM), and micro-tearing-mode (MTM) fluxes are included. The corresponding electron particle flux is $\Gamma_{e,\mathrm{total} } = \Gamma_{e,\mathrm{ETG} } + \Gamma_{e,\mathrm{ITG} } + \Gamma_{e,\mathrm{TEM} }+ \Gamma_{e,\mathrm{MTM} }$. A reduced electron particle flux model is described in \cite{Hatch2022a}. Similarly expressions could be found for ions, $q_{i,\mathrm{total} } = q_{i,\mathrm{ETG} } + q_{i,\mathrm{ITG} } + q_{i,\mathrm{TEM} }$, $\Gamma_{i,\mathrm{total} } = \Gamma_{i,\mathrm{ETG} } + \Gamma_{i,\mathrm{ITG} } + \Gamma_{i,\mathrm{TEM} }+ \Gamma_{i\mathrm{MTM} } + \Gamma_{i\mathrm{NC} }$, where neoclassical (NC) ion transport is included \cite{Helander2002}. A separate TCP would correspond to each transport channel for each species, and the most stringent of these would limit the pedestal width and height. In \Cref{fig:width_height_flow_charts}(c), we sketch how the width-height scaling might change with additional heat transport mechanisms: the upper red TCP includes just ETG transport, as in this paper. Subsequent orange and gold TCP constraints have additional electron heat transport mechanisms, which may lower the achievable pedestal pressure.

By including particle transport in addition to heat transport, one can model whether particle density builds up to the ELM limit. ELM-free transport-limited pedestals likely require either (a) an ion-scale transport mechanism that has enough particle transport to prevent density building up to the ELM limit \cite{Hatch2019} or (b) a very weak particle source \cite{Kotschenreuther2019}. One mechanism for a weak particle source is lithium evaporation causing reduced wall recycling \cite{Maingi2009}. A related important question is why some scenarios demand significant heating power but are still ELMy \cite{Hatch2019} -- strong heating sources, strong particle sources, and/or weak particle transport are a potential explanation.

Another extension potentially important is to develop a reduced model for how the particle, momentum, and heat sources vary across pedestal width and height \cite{Saarelma_2024}. We show an example of the total electron heat source $P_{e,\mathrm{limit} }$ for MAST-U 49463 in \Cref{fig:Pesource_widthheightspace}(a), and the corresponding discharge times in \Cref{fig:Pesource_widthheightspace}(b). Equilibria with higher $p_{e,\mathrm{ped} }$ values tend to have a smaller $P_{e,\mathrm{source} }$ value. This would indicate that the TCP might be even more restrictive at higher $p_{e,\mathrm{ped} }$ values than used in our model where we assumed a fixed $P_{e,\mathrm{limit} }$ across pedestal width and height. Note that for MAST-U 49483 in \Cref{fig:Pesource_widthheightspace}, the plasma triangularity was varying strongly in time \cite{Nelson2024b}, which could also have a large impact on the source. A rigorous study is required to develop reduced models of how sources vary across pedestal width and height, and the complex time-dependent coupling between transport and sources \cite{Snyder2024}.

There are several additional challenges. One is to move beyond the parameterized profiles in \Cref{eq:1,eq:2}. This requires a transport solver with accurate flux models and accurate sources. Finally, the heat flux model in \Cref{eq:qHatch23} from \cite{Hatch2024} does not explicitly contain the effects of different plasma geometry, which can to be important for ELM characteristics \cite{Merle2017,Snyder2019,Austin2019,Nelson2024,Parisi2025predictionelmfreeoperationspherical}, although recent works may include these effects through the plasma safety factor and plasma $\beta$ \cite{Guttenfelder2022,Farcas2024}.

\section{Discussion} \label{eq:discussion}

Using a threshold transport model for the electron heat flux resulting from ETG turbulence, we found Transport-Critical-Pedestal (TCP) scalings: algebraic width-height expressions for transport-limited pedestals. Initially, we analyzed electron heat transport from just a single mechanism, ETG. While none of the pedestals we analyzed from MAST-U, NSTX, or DIII-D experiments are likely limited wholly by ETG heat transport alone, some of the ELM-free pedestals are calculated to being close to ETG-limited. We also extended this model to include ETG and KBM transport through both particle and heat channels. Although we only focused on electron transport arising from ETG and KBM turbulence, the principles behind this model are generalizable to include more transport mechanisms. Given the omission of various additional transport mechanisms, the TCP in this work is approximately an upper bound on the pedestal pressure.

Based on the TCP scalings, ELM-free regimes could be achieved with techniques such as changes in the pedestal sources and an inward radial shift of the temperature pedestal compared with the density pedestal. In order for temperature-gradient-driven instability such as ETG to limit the pedestal growth, the pedestal pressure buildup usually requires some temperature contribution, otherwise the instability will not cause sufficient heat transport to limit the pedestal's growth before it hits an ELM. We also show that an additional pedestal saturation mechanism such as $\mathbf{ E} \times \mathbf{ B}$ flow shear is required if the pedestal is transport-limited rather than ELM-limited.

\section{Acknowledgements}

We are grateful for conversations with P. B. Snyder and G. M. Staebler. This work was supported by the U.S. Department of Energy under contract numbers DE-AC02-09CH11466, DE-FG02-04ER54742, DE-SC0022270, DE-SC0022272. The United States Government retains a non-exclusive, paid-up, irrevocable, world-wide license to publish or reproduce the published form of this manuscript, or allow others to do so, for United States Government purposes.

\section{Data Availability Statement}

The data used in this study will be made available on an online repository upon publication.

\appendix

\section{Power Balance} \label{app:powerbalance}

\begin{figure}[!tb]
    \centering
    \begin{subfigure}[t]{0.47\textwidth}
    \centering
    \includegraphics[width=1.0\textwidth]{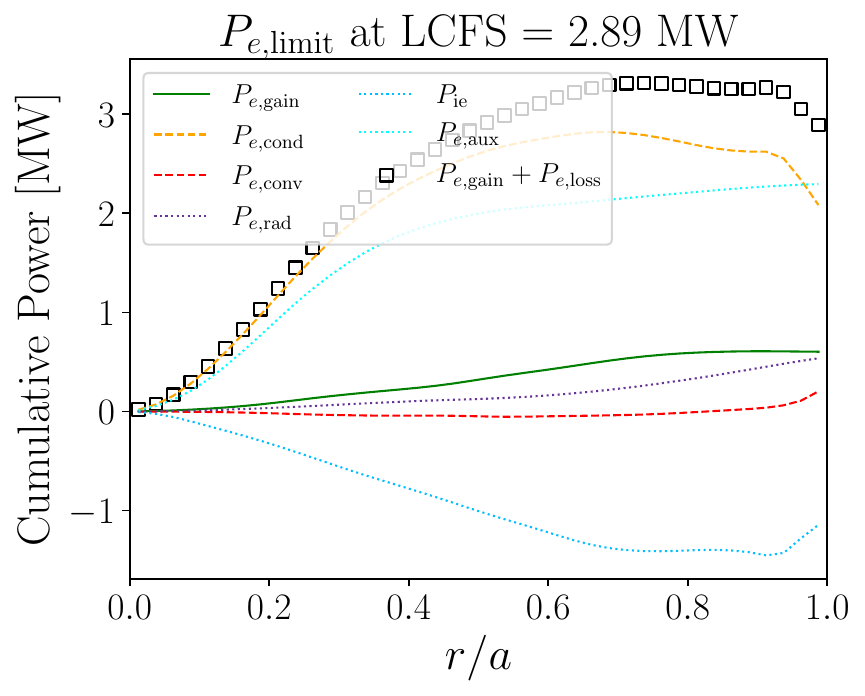}
    \caption{NSTX 132543}
    \end{subfigure}
    \begin{subfigure}[t]{0.47\textwidth}
    \centering
    \includegraphics[width=1.0\textwidth]{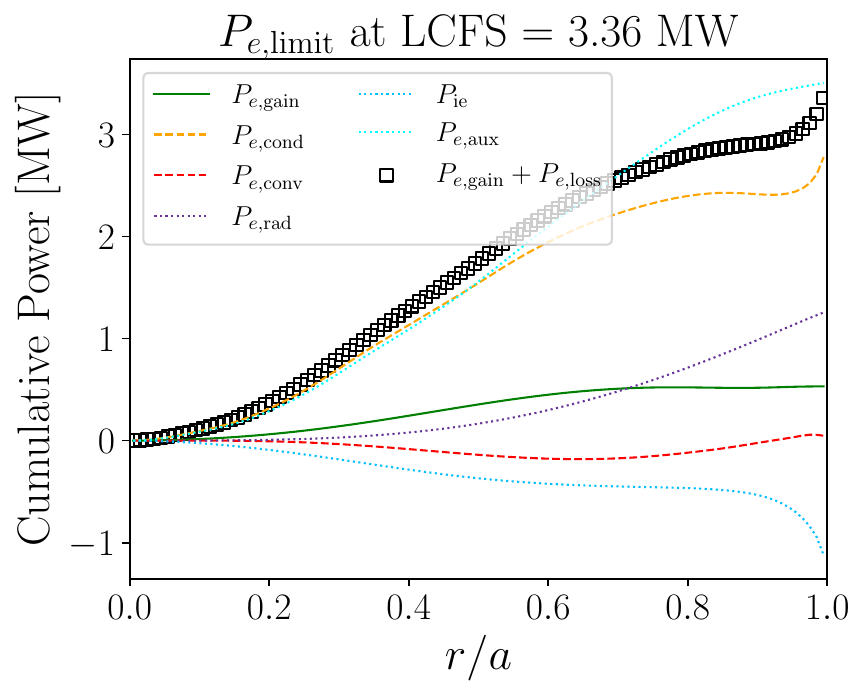}
    \caption{DIII-D 193843}
    \end{subfigure}
    \caption{Terms in the electron power balance equation, \Cref{eq:powerbalance_app}, for (a) NSTX 132543 at 614 ms and (b) DIII-D 193843 at 3800ms.}
    \label{fig:NSTX132543_sources}
\end{figure}

In this appendix, we give an example of the energy sources for an NSTX and DIII-D discharge and describe how we calculate the limiting electron source power $P_{e,\mathrm{limit} }$ from an experimental discharge. The electron energy transport equation is
\begin{equation}
\frac{3}{2} \frac{\partial  n_e T_e}{\partial t} + \nabla \cdot q_e = \sum_k S_{e,k},
\label{eq:electron_energy}
\end{equation}
which gives power balance when volume integrated from the magnetic axis up to a given flux surface,
\begin{equation}
P_{e,\mathrm{gain} } + P_{e,\mathrm{loss} } = P_{e,\mathrm{source} }.
\label{eq:powerbalance_app}
\end{equation}
The change in electron thermal energy is
\begin{equation}
P_{e,\mathrm{gain} } = \int \frac{3}{2} \frac{\partial n_e T_e}{\partial t} dV,
\end{equation}
the conductive and convective losses are
\begin{equation}
P_{e,\mathrm{loss} } = P_{e,\mathrm{cond} } + P_{e,\mathrm{conv} } = \int \nabla \cdot q_e dV,
\end{equation}
and the source terms are
\begin{equation}
P_{e,\mathrm{source} } = P_{e,\mathrm{aux} } + P_{e,\mathrm{rad} } + P_{e,\mathrm{ie} } = \int \sum_k S_{e,k} dV.
\end{equation}
Here, $P_{e,\mathrm{heat} }$ is the auxiliary electron heating power, $P_{e,\mathrm{rad} }$ is the electron radiated power, and $P_{e,\mathrm{ie} }$ is the ion-electron coupling power from electrons to ions. Because the pedestals we analyze are not steady state, $P_{e,\mathrm{gain} }$ is non-zero. Therefore, when calculating the limiting electron power for these pedestals, we assume that all of the power that goes into steepening the temperature gradient can instead be absorbed into the electron power. We set the limiting power equal to the gain and transport terms,
\begin{equation}
P_{e,\mathrm{limit}} = \left( P_{e,\mathrm{gain} } + P_{e,\mathrm{loss} } \right) \bigg{|}_{\mathrm{LCFS} }.
\label{eq:Pelimit_Pesource}
\end{equation}
In \Cref{fig:NSTX132543_sources}, we show the various electron power sources for NSTX 132543 at 614ms and DIII-D 193843 at 3800ms, predicted using the TRANSP code \cite{Pankin2024}. The limiting power $P_{e,\mathrm{limit}}$ is shown by the black squares.

\section{Particle Balance} \label{app:particlebalance}

\begin{figure}[!tb]
    \centering
    \begin{subfigure}[t]{0.47\textwidth}
    \centering
    \includegraphics[width=1.0\textwidth]{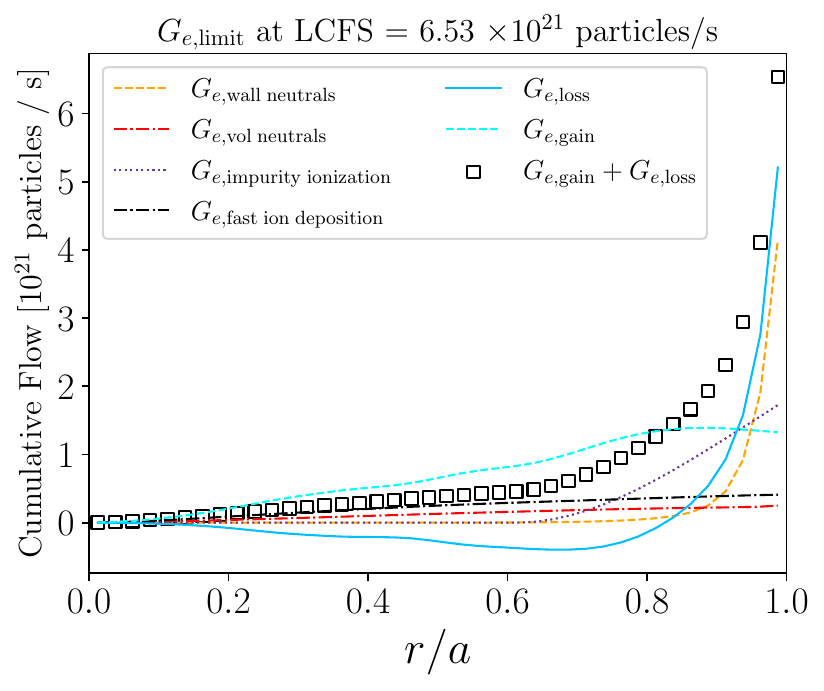}
    \caption{NSTX 132543}
    \end{subfigure}
    \begin{subfigure}[t]{0.47\textwidth}
    \centering
    \includegraphics[width=1.0\textwidth]{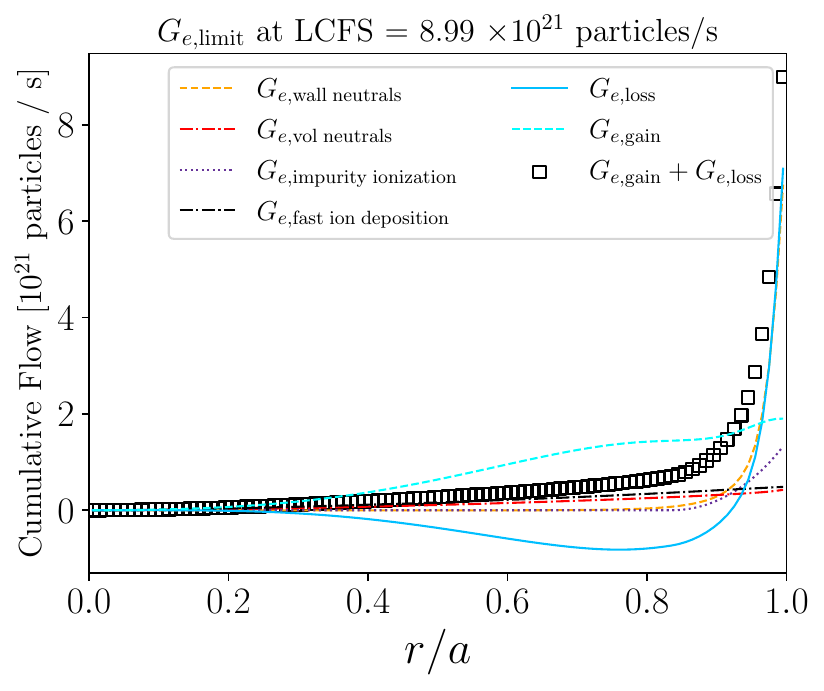}
    \caption{DIII-D 193843}
    \end{subfigure}
    \caption{Terms in the electron particle balance equation, \Cref{eq:particlebalance_app}, for (a) NSTX 132543 at 614 ms and (b) DIII-D 193843 at 3800ms.}
    \label{fig:NSTX132543_sources_particle}
\end{figure}

In this appendix, we give an example of the particle sources for an NSTX and DIII-D discharge and describe how we calculate the limiting electron particle source $G_{e,\mathrm{limit} }$ from an experimental discharge. The electron particle transport equation is
\begin{equation}
\frac{\partial n_e}{\partial t} + \nabla \cdot \Gamma_e = \sum_k L_{e,k},
\label{eq:electron_energy}
\end{equation}
which gives particle balance when volume integrated from the magnetic axis up to a given flux surface,
\begin{equation}
G_{e,\mathrm{gain} } + G_{e,\mathrm{loss} } = G_{e,\mathrm{source} }.
\label{eq:particlebalance_app}
\end{equation}
The change in electron density is
\begin{equation}
G_{e,\mathrm{gain} } = \int \frac{d n_e}{d t} dV,
\end{equation}
the loss is
\begin{equation}
G_{e,\mathrm{loss} } = \int \nabla \cdot \mathbf{\Gamma}_e dV,
\end{equation}
and the source terms are
\begin{equation}
\begin{aligned}
G_{e,\mathrm{source} } & = G_{e,\mathrm{wall \; neutrals} } + G_{e,\mathrm{vol \; neutrals} } \\
& + G_{e,\mathrm{impurity \; ionization} } + G_{e,\mathrm{fast \; ion \; deposition} } \\
& = \int \sum_k L_{e,k} dV.
\end{aligned}
\end{equation}
Here, $ G_{e,\mathrm{wall \; neutrals} }$ is the electron source from wall neutrals, $G_{e,\mathrm{vol \; neutrals} }$ is the electron source from neutrals in the main plasma, $G_{e,\mathrm{impurity \; ionization} }$ is the electron source from impurity ionization, and $G_{e,\mathrm{fast \; ion \; deposition} }$ is the electron source from fast ion deposition. Because the pedestals we analyze are not steady state, $G_{e,\mathrm{gain} }$ is non-zero. Therefore, when calculating the limiting electron particle source for these pedestals, we assume that all of the particles that go into steepening the density gradient can instead be absorbed into the electron particle flow. We set the limiting particle flow equal to the gain and transport terms,
\begin{equation}
G_{e,\mathrm{limit}} = \left( G_{e,\mathrm{gain} } + G_{e,\mathrm{loss} } \right) \bigg{|}_{\mathrm{LCFS} }.
\label{eq:Gelimit_Gesource}
\end{equation}
In \Cref{fig:NSTX132543_sources_particle}, we show the various electron particle sources for NSTX 132543 at 614ms and DIII-D 193843 at 3800ms, predicted using the TRANSP code \cite{Pankin2024}. The limiting particle flow $G_{e,\mathrm{limit}}$ is shown by the black squares.

\bibliographystyle{apsrev4-1} %
\bibliography{EverythingPlasmaBib} %

\end{document}